\documentclass[apj]{emulateapj}

\begin{document}

\title{Astrophysical Sources of Statistical Uncertainty in Precision Radial Velocities and Their Approximations}

\author{Thomas G.\ Beatty\altaffilmark{1,2} \& B.\ Scott Gaudi\altaffilmark{3}}

\altaffiltext{1}{Department of Astronomy \& Astrophysics, The Pennsylvania State University, 525 Davey Lab, University Park, PA 16802}
\altaffiltext{2}{Center for Exoplanets and Habitable Worlds, The Pennsylvania State University, 525 Davey Lab, University Park, PA 16802}
\altaffiltext{3}{Department of Astronomy, The Ohio State University, 140 W.\ 18th Ave., Columbus, OH 43210}

\shorttitle{Uncertainties in Precision Radial Velocities}
\shortauthors{Beatty \& Gaudi}

\begin{abstract}
We investigate various astrophysical contributions to the statistical uncertainty of precision radial velocity measurements of stellar spectra.  We first analytically determine the intrinsic uncertainty in centroiding isolated spectral lines broadened by Gaussian, Lorentzian, Voigt, and rotational profiles, finding that for all cases and assuming weak lines, the uncertainty is the line centroid is $\sigma_V\approx C\,\Theta^{3/2}/(W I_0^{1/2})$, where $\Theta$ is the full-width at half-maximum of the line, $W$ is the equivalent width, and $I_0$ is the continuum signal-to-noise ratio, with $C$ a constant of order unity that depends on the specific line profile.  We use this result to motivate approximate analytic expressions to the total radial velocity uncertainty for a stellar spectrum with a given photon noise, resolution, wavelength, effective temperature, surface gravity, metallicity, macroturbulence, and stellar rotation.  We use these relations to determine the dominant contributions to the statistical uncertainties in precision radial velocity measurements as a function of effective temperature and mass for main-sequence stars. For stars more massive than $\sim1.1\,M_\odot$ we find that stellar rotation dominates the velocity uncertainties for moderate and high resolution spectra ($R\gtrsim30,000$). For less massive stars, a variety of sources contribute depending on the spectral resolution and wavelength, with photon noise due to decreasing bolometric luminosity generally becoming increasingly important for low-mass stars at fixed exposure time and distance. In most cases, resolutions greater than 60,000 provide little benefit in terms of statistical precision, although higher resolutions would likely allow for better control of systematic uncertainties. We find that the spectra of cooler stars and stars with higher metallicity are intrinsically richer in velocity information, as expected. We determine the optimal wavelength range for stars of various spectral types, finding that the optimal region  depends on the stellar effective temperature, but for mid M-dwarfs and earlier the most efficient wavelength region is from 6000\,\AA\ to 9000\,\AA. 
\end{abstract}

\section{Introduction}

Current generation radial velocity (RV) surveys for exoplanets mostly focus on relatively bright, $V<8.5$, stars using single-object spectrographs. Since they are using single-object spectrographs, the surveys target one star at a time and expose up to a desired continuum signal-to-noise ratio (SNR). The $V$ cut-off used by the RV surveys is generally driven by the faint limit of the Hipparcos results \citep[e.g.,][]{marcy2005}; current RV surveys deliberately pre-select all of their stars to ensure that they will be dwarfs, chromospherically quiet, and do not have known stellar companions -- so that they will be good RV targets. Initially the surveys largely restricted their targets to single G- and K-dwarfs \citep[e.g. the stellar sample described in][]{wright2004}. This choice was driven on the one side by the faintness of M-dwarfs, which makes getting a high SNR in a reasonable exposure time difficult, and on the other side by the rapid rotation of stars hotter than the Kraft Break \citep{kraft1970} at 6250K, which widens the stellar lines and makes precision RV measurements difficult. Over the last two decades, RV surveys have thus surveyed effectively all of the single G- and K-dwarfs brighter than $V<8.5$ for Jupiter-mass planets out to periods of 5.5 years \citep[see, for example,][]{cumming2008,wright2005}.

In the last ten years, a new generation of RV surveys arose that targeted M-dwarfs \citep[M2K and the Keck M-dwarf Survey,][]{apps2010,johnson2010}, fainter high-metallicity stars \citep[N2K,][]{fischer2005}, and former A-stars that are now sub-giants \citep[``Retired A Stars,''][]{johnson2007} -- to name a few examples. These newer surveys use the same observing mode as previous surveys: single-object spectrographs and exposing to a desired SNR. They are nearly complete for Jupiter-mass planets out to several hundred days, though they have yet to survey all of the possible stars available. 

As the field moves forward, next generation of RV surveys will target fainter stars in broader observing modes. This will be enabled by the results from the GAIA mission, which will allow for vetting of target stars up to $V<20$ \citep{debruijne2012}, and also by the exhaustion of unsurveyed bright stars. There will also be a considerable demand for RV resources to follow-up planet candidates from the upcoming Transiting Exoplanet Survey Satellite \citep[TESS,][]{ricker2015} and Planetary Transits and Oscillations of stars \citep[PLATO,][]{rauer2013} missions, as well as residual planet candidates from the Kepler \citep{borucki2010} mission. Importantly, these photometric exoplanet surveys do not pre-select their targets in the same manner as the traditional RV surveys (if they pre-select at all). Vetting the candidates from these missions will therefore necessitate precision RV measurements of stars that are potentially more active, hotter, more evolved, and rotating faster than the targets of the first-generation surveys.
 
In all of this, understanding the sources of velocity uncertainty in stellar RV measurements is critical. This allows us to both appropriately understand the detection sensitivities of current RV surveys, and more efficiently design and execute future precision RV searches for exoplanets. In particular, consider a multi-object precision RV survey, which would have design considerations very different from present single-star searches. Firstly, a multi-object survey will image a much narrower spectral region than a single-object survey at a similar resolution, on the order of 100\,\AA\ for a multi-object \citep[e.g.,][]{furesz2008} versus 1500\,\AA\ for a single object survey \citep[e.g.,][]{marcy1996}. This has the immediate implication that choosing the wavelength range used by a multi-object survey will be much more important than in a single-object survey. Secondly, a multi-object RV survey may not conduct an extensive pre-survey vetting of targets. This will give their target lists a higher dispersion in stellar mass, rotation velocities, activity, surface gravity, and metallicity than current single-object surveys. Finally, multi-object surveys will operate in an observing mode more similar to photometric transit surveys for exoplanets, as compared to traditional RV surveys. Instead of exposing to reach the same SNR for each of the survey's target stars, a multi-object survey will expose for the same time on each star. This will make the detection sensitivities of a multi-object fundamentally different from a single-object survey in terms of stellar mass, effective temperature, and metallicity. A proper understanding of how these sensitivities change is vital to the initial design of a multi-object survey, and understanding the exoplanet statistics from the survey when it has completed.

There have been several efforts to study the sources of radial velocity uncertainty since the beginnings of precision RV surveys. The first description of how to numerically calculate the expected velocity uncertainty using the properties of an observed spectrum was given by \cite{connes1985}, and who also introduced the ``Q'' factor as a way of parameterizing the amount of RV information available in a spectral region, though they do not discuss how this is expected to vary with spectral properties. \cite{murdoch1991} performed a similar analysis, but started from the idea of measuring cross-correlation functions, rather than measuring spectral slopes as in \cite{connes1985}, which makes for an interesting conceptual comparison between these two works. The approach used by \cite{connes1985} was repeated in \cite{butler1996}, who were interested in comparing theoretical Poisson-limited velocity uncertainties against their actual observations. Similar to \cite{connes1985}, they did not discuss how the uncertainty varies with stellar properties. \cite{bouchy2001}, after presenting their own derivation of the photon-limited RV precision, were the first to examine how velocity uncertainty changes as a function of spectral type, stellar rotation velocity, and spectroscopic resolution. Again, these results were used as part of a general discussion of the capabilities of the CORALIE \citep{queloz2000} and the (then future) HARPS \citep{mayor2003} spectrographs. \cite{bouchy2001} thus concentrated their analysis on a limited range of spectral types, rotations, and resolutions, and drew descriptive conclusions from their results (for example, they inferred that uncertainty is proportional to $v \sin i$ when $v \sin i$ is large). More recently, \cite{bottom2013} considered the sources of uncertainty in RV measurements as part of an examination into how to optimize RV surveys of GKM stars. Unlike \cite{butler1996} and \cite{bouchy2001}, \cite{bottom2013} did not derive an equation for the photon-limited uncertainty in an RV observation, but took model spectra, added Gaussian noise, and then fit for the Doppler shift in the noisy spectra using cross-correlation techniques. The authors intent was to better replicate the true process behind the measurement of stellar RVs. \cite{bottom2013} consider wider and more finely-spaced ranges of wavelength, temperature, and spectral resolution compared to \cite{bouchy2001}, but they do not consider in detail the effect of changing stellar rotation. Similarly to \cite{bouchy2001}, \cite{bottom2013} restrict themselves to descriptive conclusions regarding the sources of velocity uncertainty.

In this paper, we aim to provide a more thorough description of photon-limited stellar velocity uncertainties. We do not consider the effects of ``jitter'' sources such as spots, granulation, or asteroseismic pulsations. Similarly, we do not treat instrumental velocity uncertainties like wavelength calibration, optical effects, or instrumental drifts, nor do we consider the effect of telluric absorption lines on the spectra. Instead, we are here focused on the statistical velocity uncertainty one would measure for a photospherically stable rotating star using a perfect instrument through a completely transparent atmosphere.  

Starting from a basic derivation of how to calculate the velocity uncertainty in a spectrum, our intent is to consider, in detail, the effects of effective temperature, surface gravity, metallicity, stellar rotation, spectral resolution, and macroturbulence on the uncertainty in an RV measurement. This allows us to make not only descriptive, but prescriptive conclusions regarding the sources of uncertainty. For example, we discuss precisely why stellar rotation has the effect it does on velocity uncertainties, starting from the shape of the rotation kernel itself. We are thus able to simply and numerically describe how rotation affects the velocity uncertainty across all rotation velocities.

Our ultimate goal is to provide a simple and transparent description of the various sources of velocity uncertainty in RV measurements, so as to give the reader a clear picture of the interlocking forces at work. To this end, we use our results to provide several simplifying approximations that capture the dominant sources of velocity uncertainty as function of stellar mass. This allows us to better understand why current RV surveys achieve the precision they do, and to provide guidance for the design of future surveys. 

\section{Photon-limited Radial Velocity Precision}

Consider a one-dimensional spectrum $I_i$ that is discretely sampled at equal velocity intervals (e.g., bins or pixels) $\Delta V$ centered at velocities $V_i$.  The number of photons in each velocity bin $\Delta V$ is $N_i=I_i\Delta V$.  We will assume for simplicity that the spectrum is in units of photons per pixel or per resolution element, and the uncertainties are Poisson-dominated so that the uncertainty in $N_i$ is just $\sigma_{N,i}=\sqrt{N_i}$.  We will further assume that these uncertainties are uncorrelated. 

Although the precise details vary depending on the method and instrumental setup, we can think of the process of determining the radial velocity of this spectrum as a cross-correlation of a template spectrum of the star (either from models or a previous observation) against the observed spectrum. Typically, the peak or centroid of the cross-correlation function output by the spectral template matching is assumed to be the best estimate of radial velocity of the spectrum relative to the template. We wish to determine what the uncertainty is in this estimate. As described in the Introduction, this general problem has been considered before, and the following derivation of the general answer is similar to previous work.

As mentioned, we will assume that the uncertainties in our imagined spectrum are purely from Poisson noise, are free of any systematics, and uncorrelated. Furthermore, we will assume that the velocity offset is not covariant with any other possible fitting parameters. We are therefore, in some sense, determining the best possible precision on the velocity offset that can be obtained, given the structure of the spectrum and assuming only source photon noise.  There are various ways to compute the lowest possible uncertainty in the velocity offset \citep[i.e., the minimum variance bound,][]{gould1995}, for example using a Fisher matrix estimation. However, we will simply use the maximum likelihood estimator and error propagation, which is mathematically equivalent to the Fisher estimate under our set of assumptions.

We will assume that each discrete sampling of the spectrum yields an estimate of the velocity $V_i$ of the spectrum with an uncertainty $\sigma_i$, and the best estimate of the velocity $V$ of the entire spectrum is the weighted mean over all the samples,
\begin{equation}\label{eq:2020}
V = \frac{\sum_i w_i V_i}{\sum_i w_i},
\end{equation}
where $w_i = 1/\sigma_i$.  The variance in this mean is then determined from the general error propagation equation, i.e., by differentiating the maximum-likelihood function and keeping only the first order terms, such that,
\begin{eqnarray}\label{eq:2010}
\sigma_V^2 &=& \sigma_1^2\left(\frac{\partial V}{\partial V_1}\right)^2 + \sigma_2^2\left(\frac{\partial V}{\partial V_2}\right)^2 + ... \\ \nonumber
&=& \sum_i \left[\sigma_i^2\left(\frac{\partial V}{\partial V_i}\right)^2\right].
\end{eqnarray}
From Equation (\ref{eq:2020}) the partial derivatives in Equation (\ref{eq:2010}) are then
\begin{equation}\label{eq:2030}
\frac{\partial V}{\partial V_i} = \frac{w_i}{\sum_i w_i} = \frac{1/\sigma_i^2}{\sum_i 1/\sigma_i^2}.
\end{equation}
Substituting and simplifying, we obtain \citep[see also][]{butler1996},
\begin{equation}\label{eq:2040}
\sigma_V^2 = \left(\smash{\sum_i} \frac{1}{\sigma_i^2} \right)^{-1}.
\end{equation} 

We can relate the uncertainty in the velocity inferred for each point in the spectrum $\sigma_i$ to the uncertainty in the intensity (photon number) at that point $\sigma_{I,i}$ via the local derivative of the spectrum with velocity,
\begin{equation}\label{eq:2050}
\sigma_i = \frac{\sigma_{N,i}}{(dN/dV)|_i}.
\end{equation}
Again assuming through Poisson statistics that $\sigma_{N,i}=\sqrt{N_i}$, and subsisting the above into Equation
(\ref{eq:2040})
\begin{equation}\label{eq:2060}
\sigma_V^2 = \left[\sum_i \frac{(dN/dV)^2|_i}{N_i} \right]^{-1}.
\end{equation}
This is generally how well we can measure the velocity positions of the features in an arbitrary function of intensity versus velocity. This is similar to the ``Q'' factor formulation first used stated in \cite{connes1985}.

\subsection{Centroiding absorption lines}

\begin{figure}
\vskip -0.0in 
\epsscale{1.2} 
\plotone{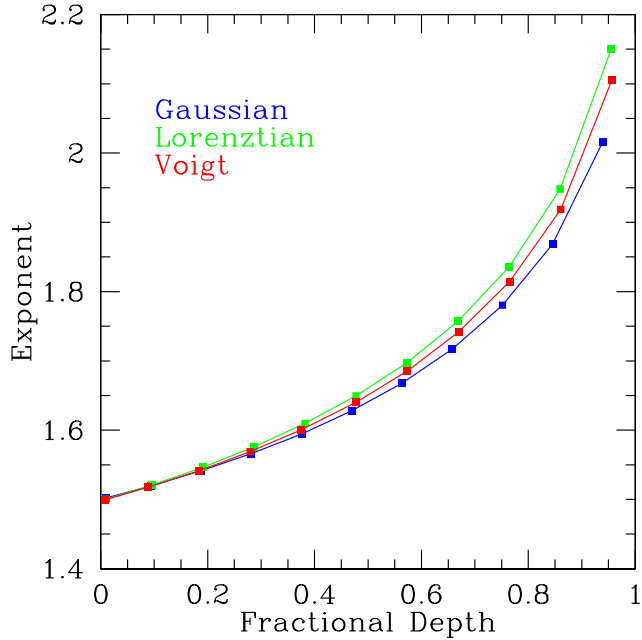}
\vskip -0.0in 
\caption{The exponent with which the velocity uncertainty $\sigma_V$ scales with the width $\Theta$ of an absorption line, i.e., $\sigma_V\propto \Theta^n$, as a function of the line depth relative to the continuum level. This is numerically calculated using Equation (\ref{eq:2060}) for each of the three line profiles. At a fractional depth of $\approx0$ we find that the uncertainty scales as $\sigma_V\propto \Theta^{1.5}$, which is the same as for our analytic approximations for shallow lines given in Equations (\ref{eq:2160}), (\ref{eq:2170}), and (\ref{eq:2190}).}
\label{fig:deep}
\end{figure}

Let us first consider the case of a Gaussian absorption line. For compactness, let us define $G(V_i,V_0,\Theta_G)$ as the appropriately normalized Gaussian distribution centered at $V_0$ and with a full-width at half-max (FWHM) $\Theta_G$: 
\begin{equation}\label{eq:2105}
G(V_i,V_0,\Theta_G) = \sqrt{\frac{4\ln2}{\pi \Theta_G^2}} \exp\left[\frac{-(V_i-V_0)^2}{\Theta_G^2/(4\ln2)}\right].
\end{equation}
If we take a spectrum composed of points separated by a constant $\Delta V$ in velocity, and if this line absorbs $N_{tot}$ photons, then a spectrum containing only this line can be described by
\begin{equation}\label{eq:2110}
N_\gamma(V_i) = (I_0 - N_{tot} G[V_i,V_0,\Theta_G]) \Delta V,
\end{equation}
where $I_0$ is the continuum level, in units of photons per unit velocity, and $N_\gamma(V_i)$ is the number of photons in a particular velocity bin centered at $V_i$. The factor of $\Delta V$ is the velocity span of a pixel. We can rewrite this equation in terms of the velocity equivalent width of the line, $W\equiv N_{tot}/I_0$, as
\begin{equation}\label{eq:2120}
N_\gamma(V_i) = I_0 \Delta V (1-WG) = N_{\gamma,\mathrm{cont}}(1-WG).
\end{equation}
Substituting into Equation (\ref{eq:2060}), we get
\begin{equation}\label{eq:2130}
\sigma_V^2 = \left[\sum_i (V_i-V_0)^2 \left(\frac{2\sqrt{2\ln2}}{\Theta_G}\right)^4 \frac{I_0^2 \Delta V^2 W^2 G^2}{I_0\Delta V(1-WG)} \right]^{-1}.
\end{equation}
By the Euler-Maclaurin formula, we can approximate this sum with the integral
\begin{eqnarray}\label{eq:2140}
\sigma_V^2 = \biggr[\frac{1}{\Delta V} \int_{-\infty}^\infty (V-V_0)^2 && \left(\frac{2\sqrt{2\ln2}}{\Theta_G}\right)^4 \\ \nonumber
&& \frac{I_0^2 \Delta V^2 W^2 G^2}{I_0\Delta V(1-WG)} dV \biggr]^{-1}.
\end{eqnarray}
Unfortunately, this integral has no analytic solution. As a limiting case, consider a shallow absorption line such that $1-WG\approx1$. Now we may analytically solve the above equation to get
\begin{equation}\label{eq:2150}
\sigma_V^2 = \left[\frac{I_0 W^2}{4\sqrt{\pi}} \left(\frac{2\sqrt{2\ln2}}{\Theta_G}\right)^3 \right]^{-1},
\end{equation}
or,
\begin{equation}\label{eq:2160}
\sigma_V = \left(\frac{\sqrt{\pi}}{2(2\ln2)^{3/4}}\right) \frac{\Theta_G^{3/2}}{W\sqrt{I_0}} \approx 0.69\, \frac{\Theta_G^{3/2}}{W\sqrt{I_0}}.
\end{equation}
In numerical tests (Figure \ref{fig:deep}), we find that this approximation is valid for lines with depths less than about 10\% of the continuum level. For lines deeper than this the exponential dependence on the width of the Gaussian increases. By the time the depth of the line is 95\% of the continuum level we find that the uncertainty in the centroid scales roughly as $\Theta_G^2$.   

The uncertainty in the centroid of a shallow Gaussian absorption line therefore scales approximately as the FWHM of the line to the three-halves \citep[see also][]{lovis2011}. We can understand this intuitively by approximating an absorption line as triangular in shape, with a peak that is lower than the continuum by a fractional amount $1/\Theta$, and a width of $\Theta$. The form for such a line is simply:
\begin{equation}\label{eq:2165}
N_\gamma(V_i) = I_0 \Delta V \left(1- W\left|\frac{V_i-V_0-\Theta}{\Theta^2}\right|\right),
\end{equation}
for $|V_i-V_0|<\Theta$ and $N_\gamma(V_i)=I_0\Delta V$ otherwise.  For a weak line, the second term in parenthesis in Equation (\ref{eq:2165}) is negligible, and applying Equation (\ref{eq:2060}) yields
\begin{equation}\label{eq:2167}
\sigma_V = \frac{1}{\sqrt{2}}\frac{\Theta^{3/2}}{W\sqrt{I_0}} \approx 0.71\, \frac{\Theta^{3/2}}{W\sqrt{I_0}},
\end{equation}
with the same scaling and very similar coefficient as the weak Gaussian absorption line case.  Conceptually, we can understand this as follows.  The amount of information decreases, and the uncertainty increases, as the inverse of slope of the triangular line: $\sigma_V \propto \Theta^2$.  On the other hand, the amount of information increases, and the uncertainty decreases, as the square-root of the total number of photons. In the approximation of a weak line this is proportional to the square-root of the velocity span of the continuum covered by the base of the triangular line, and thus $\sigma_V\propto \Theta^{-1/2}$. The net result of these two competing effects is that $\sigma_V \propto \theta^{3/2}$.  This general argument implies that uncertainty in the centroid for all weak lines in a background continuum, regardless of their detailed profile, should scale as $\Theta^{3/2}$.  We demonstrate this explicitly for the other velocity profiles we consider below.  We note that this scaling contrasts with the uncertainty of measuring the centroid of a pure Gaussian profile with zero continuum (i.e., calculating the standard error of the mean), which has the scaling of \citep[e.g.,][]{kenney1947}
\begin{equation}\label{eq:2168}
\sigma_V = \frac{\sigma}{\sqrt{n}} = \frac{1}{2\sqrt{2\ln 2}} \frac{\Theta}{\sqrt{I_0 W}}, 
\end{equation}
where $\sigma$ is the standard deviation of the Gaussian profile, and $n$ is the number of observations in the sample.

Similarly to the Gaussian, we can work through the corresponding derivation for a Lorentzian absorption line with FWHM $\Theta_L$ to find
\begin{equation}\label{eq:2170}
\sigma_V = \sqrt{\frac{\pi}{2}}\ \frac{\Theta_L^{3/2}}{W\sqrt{I_0}} \approx 1.25\, \frac{\Theta_L^{3/2}}{W\sqrt{I_0}}.
\end{equation}
This analytic solution is, again, under the assumption that the line depth is negligible relative to the continuum level. As can be seen in Figure \ref{fig:deep}, the exponent on $\Theta_L$ increases as the line depth increases, similar to the Gaussian case.

Often, spectral lines are effectively described with Voigt profiles: the convolution of a Lorentzian and a Gaussian. Though there is no analytic description of a Voigt profile, we numerically examined how the uncertainty in the centroid scales with the Voigt width. We used \cite{olivero1977}'s approximation for the effective FWHM of a Voigt profile,
\begin{equation}\label{eq:2180}
\Theta_{\mathrm{Voigt}} = 0.5346\Theta_L + \sqrt{0.2166\Theta_L^2 + \Theta_G^2},
\end{equation}
where $\Theta_L$ is the FWHM of the Lorentzian component and $\Theta_G$ is the Gaussian FWHM. We directly calculated the uncertainty in the centroid using Equation (\ref{eq:2060}) for various values of $\Theta_L$ and $\Theta_G$. As with the pure Gaussian and pure Lorentzian, we find that the uncertainty in measuring the centroid of a Voigt profile is proportional to $\Theta_{\mathrm{Voigt}}^{3/2}$. For the constant of proportionality relating $\sigma_V$ and $\Theta_{\mathrm{Voigt}}$, we found numerically that it varied as a function of the ratio $\Theta_L/\Theta_G$.
\begin{equation}\label{eq:2190}
\sigma_V = 0.96\left(\frac{\Theta_L}{\Theta_G}\right)^{1/2}\ \frac{\Theta_{\mathrm{Voigt}}^{3/2}}{W\sqrt{I_0}}.
\end{equation}
This form of the leading constant is good to 10\% over the range of $1/5<\Theta_L/\Theta_G<5$.

In addition to spectral absorption lines, we will also need to consider the role stellar rotation plays in setting measured RV uncertainties. We begin by considering the shape of the kernel itself to determine the FWHM of the rotation kernel: $\Theta_{rot}$. To do so, we must first assume a limb-darkening law. For simplicity, we use a simple linear limb-darkening law: $I=I_0(1-\epsilon+\epsilon\cos\theta)$, where $I_0$ is the intensity at the center of the stellar disk, $\theta$ is the angle of the surface to our line of sight, and $\epsilon$ is the limb-darkening coefficient. Following \cite{gray2008}, the normalized rotation kernel is then
\begin{eqnarray}\label{eq:3165}
G(\Delta v) &=& \frac{2(1-\epsilon)\sqrt{1-(\Delta v/v_{rot})^2}}{\pi v_{rot}(1-\epsilon/3)} \\ \nonumber
&& + \frac{\frac{1}{2}\pi\epsilon(1-(\Delta v/v_{rot})^2)}{\pi v_{rot}(1-\epsilon/3)},
\end{eqnarray}
where $v_{rot}$ is the rotation speed at the limb of the star.  For the case of no limb-darkening ($\epsilon=0$) this reduces to 
\begin{equation}\label{eq:3156}
G(\Delta v) = \frac{2}{\pi v_{rot}}\,\sqrt{1-(\Delta v/v_{rot})^2}.
\end{equation}
To find $\Theta_{rot}$ we then set $G(\Delta v)=1/\pi v_{rot}$ (i.e., half the maximum), $\Delta v=\Theta_{rot}/2$, and solve. Thus         
\begin{equation}\label{eq:3160}
\Theta_{rot} = \sqrt{3}\ v_{rot}\ \ (\mathrm{for\ \epsilon=0}).
\end{equation}
At the other extreme of $\epsilon=1$, we may solve Equation (\ref{eq:3165}) to find 
\begin{equation}\label{eq:3167}
\Theta_{rot} = \sqrt{2}\ v_{rot}\ \ (\mathrm{for\ \epsilon=1}).
\end{equation}
Aside for the cases of $\epsilon=0$ and $\epsilon=1$, Equation (\ref{eq:3165}) allows for no simple analytic formula for $\Theta_{rot}$ as a function of $v_{rot}$ and $\epsilon$. We therefore numerically measured the FWHM of several calculated kernels between $0<\epsilon<1$. We found that in between the two limb-darkening extremes the FWHM went linearly with $\epsilon$, such that
\begin{equation}\label{eq:3168}
\Theta_{rot} = [(\sqrt{2}-\sqrt{3})\,\epsilon + \sqrt{3}]\ v_{rot}.
\end{equation}
This relation is accurate to better than 5\% over the entire range of $0\leq\epsilon\leq1$. For reference, a Sun-like star observed at 5500\,\AA\ would have $\epsilon\approx0.75$, and $\epsilon\approx0.4$ if observed at 10,000\,\AA.

Now, similar to the absorption line profiles, we may use Equation (\ref{eq:2060}) to determine how velocity uncertainty scales with the width of the rotation kernel. We first consider the case of a fully limb-darkened kernel with $\epsilon=1$ and equivalent width $W$ subtracted from a continuum. Thus the spectrum is given by $N_\gamma(V_i) = N_{\gamma,\mathrm{cont}}(1-WG)$. This represents the ideal case of a $\delta$-function absorption line being rotationally broadened by the kernel. After making the appropriate substitutions into Equation (\ref{eq:2060}), and again assuming that $1-WG\approx1$, we find that
\begin{equation}\label{eq:3190}
\sigma_V = \sqrt{\frac{2/3}{2^{3/2}}}\,\frac{\Theta_{rot}^{3/2}}{W\sqrt{I_0}} \approx 0.49\,\frac{\Theta_{rot}^{3/2}}{W\sqrt{I_0}}\ \ (\mathrm{for\ \epsilon=1}).
\end{equation}

\begin{figure}[t]
\vskip -0.0in 
\epsscale{1.2} 
\plotone{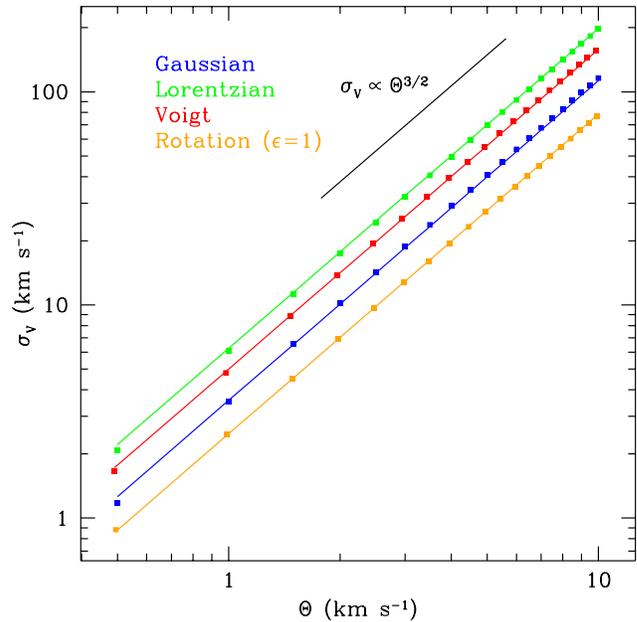}
\vskip -0.0in 
\caption{The colored points show how the velocity uncertainty varies with FWHM for absorption lines with a Gaussian, Lorentzian, Voigt, or rotational profile with $W=0.2$\,km s$^{-1}$ and $I_0=1$ photon per km s$^{-1}$, determined via direct numerical computation using Equation (\ref{eq:2060}). The overplotted lines are the directly calculated velocity uncertainties for these same absorption lines using our analytic approximations in Equations (\ref{eq:2160}), (\ref{eq:2170}), (\ref{eq:2190}), and (\ref{eq:3190}). For the Voigt profile we set the Lorenztian and Gaussian widths equal, so that $\Theta_L/\Theta_G=1$ in Equation (\ref{eq:2190}).}
\label{fig:singleline}
\end{figure}

Unfortunately, $\epsilon=1$ is the only case for which we may calculate $\sigma_V$ directly from the rotation kernel itself using Equation (\ref{eq:2060}). For all other values of $\epsilon$ the slope of the kernel goes to infinity as $\Delta V\to v_{rot}$. We were not able to find an appropriate analytic or numeric integral to avoid this, so we instead convolved kernels for $\epsilon<1$ with a normalized Gaussian of small fixed width ($\sigma=0.1$ km s$^{-1}$) and measured the velocity uncertainties of the resulting lines for $v_{rot}>25$ km s$^{-1}$. We found that the velocity uncertainty continued to be proportional to $\Theta_{rot}^{3/2}$, with the constant of proportionality varying with roughly linearly $\epsilon$, such that 
\begin{equation}\label{eq:3192}
\sigma_V \approx (0.347+0.146\,\epsilon)\,\frac{\Theta_{rot}^{3/2}}{W\sqrt{I_0}}
\end{equation}   
This is accurate to 2\% over $0\leq\epsilon\leq1$. 

Interestingly, if one rewrites Equation (\ref{eq:3192}) in terms of $v_{rot}$, rather than $\Theta_{rot}$, the $\epsilon$ dependence of $\sigma_V$ nearly cancels out. Put another way, velocity uncertainties are not strongly effected by the precise amount of limb-darkening in the stellar photosphere. The difference in $\sigma_V$ between $\epsilon=0$ and $\epsilon=0.75$ (the locations of the minimum and maximum of the proportionality coefficient) is only 5\%. This is somewhat dependent on our choice of a linear limb-darkening law, but this result should approximately hold for more complicated limb-darkening laws.

\begin{figure}[b]
\vskip -0.0in 
\epsscale{1.2} 
\plotone{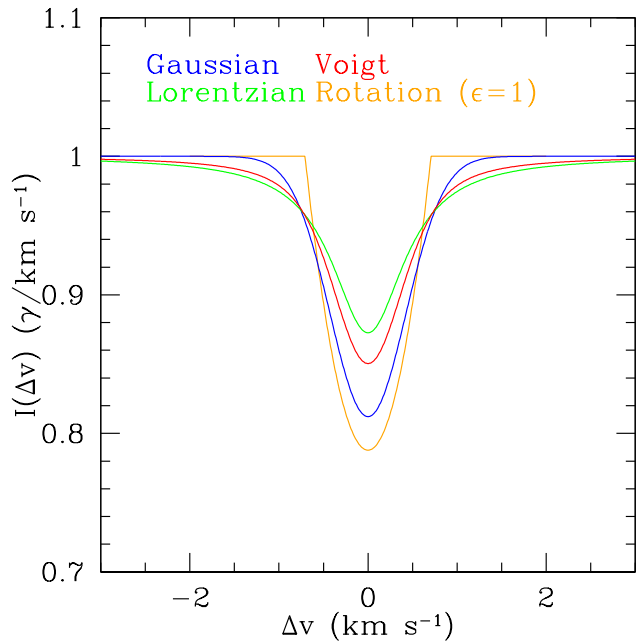}
\vskip -0.0in 
\caption{The different line broadening profiles we consider in Section 2.1, all with the same FWHM ($\Theta=1$\,km s$^{-1}$) and equivalent width ($W=0.2$\,km s$^{-1}$).}
\label{fig:profiles}
\end{figure}

Figure \ref{fig:singleline} shows the directly calculated uncertainties for a rotation kernel, Gaussian, Lorentzian, and Voigt profile with the same equivalent width as a function of FWHM. The overplotted lines are what we expect for the uncertainty based on Equations (\ref{eq:2160}), (\ref{eq:2170}), and (\ref{eq:2190}). For reference, Figure \ref{fig:profiles} shows all four profiles, each one with $\Theta=1$ km s$^{-1}$ and $W=0.2$ km s$^{-1}$.

\section{Calculating Stellar Velocity Uncertainties}

Our ultimate goal is to calculate the photon-limited uncertainty expected for RV observations of a main sequence star with a certain exposure time. To achieve this, we will examine how the velocity uncertainties of synthetic spectra -- calculated directly using Equation (\ref{eq:2060}) -- change as a function of instrumental and stellar properties. So far we have been describing the behavior of individual lines; going forward we will instead consider the average behavior of groups of lines over an extended spectrum.

From our consideration of the uncertainty in centroiding single absorption lines, we anticipate that the velocity uncertainty of a group of lines should scale as
\begin{equation}\label{eq:3010}
\sigma_V \propto \sqrt{N_{\mathrm{lines}}}\ \frac{\bar{\Theta}^{3/2}}{W\sqrt{I_0}},
\end{equation}
where $I_0$ is the continuum flux level, $\bar{\Theta}$ and $\bar{W}$ are the average FWHM and equivalent widths of the lines in the spectrum of interest, and $N_{\mathrm{lines}}$ is the number of lines in the spectrum. The continuum level is set by the spectral energy distribution of the target star, the exposure time, the size of the telescope, and the throughput of the optical system. The average FWHM of the spectral lines is determined by the spectral resolution of the instrument, the observed wavelength range, as well as a host of stellar parameters such as mass, temperature, age, rotational velocity and metallicity. We chose to restrict ourselves to stars that are on the main sequence, specifically dwarfs from 2600K to 7600K. This allows us to parameterize stellar properties in terms of one parameter (e.g., mass or effective temperature). This leaves us with seven main parameters that will determine the average FWHM of the spectral lines: spectral resolution, wavelength range, stellar rotation, photospheric macroturbulence, effective temperature, surface gravity, and metallicity.

For simplicity, we chose to quantify $N_{\mathrm{lines}}$, the number of lines present in a spectrum, and $\bar{W}$, their average equivalent widths, using separate and purely descriptive functions of effective temperature, surface gravity, and metallicity. Thus Equation (\ref{eq:3010}) becomes
\begin{equation}\label{eq:3020}
\sigma_V \propto \frac{\bar{\Theta}^{3/2}}{\sqrt{I_0}}\ f(T_{eff})\ f(\log g)\ f(\mathrm{[Fe/H]}).
\end{equation}
This is the approximation we will use in our fitting.

We are making an important assumption here: that $\bar{\Theta}$ exists as a meaningful descriptor for a given spectrum. We intend to calculate $\bar{\Theta}$ for a spectrum as a combination of some $\Theta_0$ caused by the inherent widths of the stellar lines, and an additional line width caused by one or more broadening mechanisms (e.g., $\Theta_{R}$ for spectral resolution). The assumption we are making therefore allows us to determine $\Theta_0$ not by measuring the FWHM of each line in a given spectrum, but rather by fitting how that spectrum's velocity precision scales as a source of broadening is applied to it. If this assumption that $\bar{\Theta}$ exists as a meaningful quantity is correct, then we expect that $\sigma_V$ for spectra of varying wavelengths will behave roughly self-similarly as we apply various broadening mechanisms, and that spectra over a wide wavelength range will have fitted values of $\Theta_0$ that are similar. For the moment, let us accept this underlying assumption as correct. We shall see in subsequent subsections that $\sigma_V$ does indeed behave self-similarly, and that a single value of $\Theta_0$ can describe a spectrum's response to broadening for different spectral regions spaced over thousands of angstroms.  

We used two different sets of synthetic spectra in our fitting: one set from the BT-Settl model spectra \citep{allard2012} and another calculated using Kurucz model atmospheres \citep[][hereafter Kurucz92]{kurucz1992}. The two sets provide us with different pieces of information: the BT-Settl spectra cover the full temperature range of interest, but have a (relatively) coarse wavelength spacing, while the Kurucz92 spectra are extremely finely spaced in wavelength but are only available for a subset of the temperatures in which are interested. We therefore used the BT-Settl spectra to examine the effect of stellar effective temperature and surface gravity, and used the Kurucz92 spectra to model line broadening mechanisms like spectral resolution and stellar rotation. Since one can generically think of these latter mechanisms as externalities imposed upon ``perfect'' stellar spectra (i.e., rotational broadening is not an intrinsic part of creating absorption lines) we expected the results we find using the Kurucz92 spectra to be consistent with our results using the BT-Settl spectra, once we correct for the difference in wavelength sampling. As described later, we ultimately found this to be the case.

We used flux-normalized spectra in three broad bands: ``optical'' spectra from 4000\,\AA\ to 6500\,\AA, ``red'' spectra from 6500\,\AA\ to 10000\,\AA, and ``near-infrared'' (NIR) spectra from 10000\,\AA\ to 25000\,\AA. The BT-Settl spectra used the \cite{asplund2009} solar abundances, had solar metallicity with no $\alpha$-enhancement, and were spaced 200K apart from 2600K to 7600K. The BT-Settle spectra available for download use a variable wavelength spacing, with a finer spacing occurring around the absorption lines. On average the wavelength spacing was 0.05\,\AA\ ($R\approx100,000$) in the optical, 0.05\,\AA\ ($R\approx165,000$) in the red, and 0.2\,\AA\ ($R\approx88,000$) in the NIR.

For the Kurucz92 spectra we used the {\sc odfnew} versions of the Kurucz92 model atmospheres with no $\alpha$-enhancement, and generated the spectra with v2.76 of \cite{gray1994}'s {\sc spectrum} code. For the optical and red spectra we used a fixed wavelength spacing of 0.001\,\AA\ ($R\approx6\times10^6$) and for the NIR spectra we used 0.005\,\AA\ ($R\approx3.5\times10^6$). In all the bands we set the microturbulent velocity to 1 km s$^{-1}$ and left the macroturbulent velocity at zero. We considered the effect of macroturbulence separately. The Kurucz92 models covered effective temperatures from 4000K to 7500K with a spacing of 250K. 

To investigate wavelength dependent features, we divided all our spectra into 100\,\AA\ chunks. This partially isolates individual line groupings, like the Mg B triplet, so that we can test whether these groupings react to changes in a self-similar way. Furthermore, splitting the spectra into 100\,\AA\ chunks allowed us to mirror the actual analysis procedures of current multi-order RV surveys, and is representative of the amount of spectral information available in proposed multi-object surveys. 

For each chunk, we calculated the expected velocity uncertainty using Equation (\ref{eq:2040}). After transforming each chunk from wavelength to velocity space, we normalized the chunks so that each had $N_i=1$ photon per m s$^{-1}$ in the continuum ($I_0=1$). We then numerically calculated the slope of the spectrum at each pixel ($(dN/dV)|_i$). By Equation (\ref{eq:2040}), this then gives us a velocity uncertainty for each 100\,\AA\ chunk. 

We now wish to see how the velocity uncertainties calculated from the individual 100\,\AA\ chunks vary as we vary spectral resolution, stellar rotation, macroturbulence, effective temperature, surface gravity, and metallicity.
 
\subsection{Spectral Resolution, Stellar Rotation, and Macroturbulence: Kurucz92 Based}

Conceptually, the effect on a spectrum of changing spectral resolution ($R$), stellar rotation ($v\sin i$) or the macroturbulent velocity ($v_{mac}$) can be viewed as an externality imposed upon a ``perfect'' spectrum with $R=\infty$, $v\sin i$=0 and $v_{mac}$=0. Regardless of the underlying stellar parameters, the velocity uncertainty of a spectrum should vary (roughly) with the same functional form for $R$, $v\sin i$ and $v_{mac}$. We therefore rewrite Equation (\ref{eq:3010}) to 
\begin{equation}\label{eq:3110}
\sigma_V = \sigma_{V,0}\,[\varphi_{rel}(R,v_{rot},v_{mac})]^{3/2},
\end{equation}
where $\sigma_{V,0}$ is the velocity uncertainty of a ``perfect'' spectrum with $R$=$\infty$, and no rotation or macroturbulence. $\varphi_{rel}(R,v_{rot},v_{mac})$ is the increase in the average FWHM of the spectral lines caused by changes in $R$, $v\sin i$, and macroturbulence relative to that of the perfect spectrum. We defined $\varphi_{rel}$ such that $\varphi_{rel}(\infty,0,0)$=1 and $\varphi_{rel}(0,\infty,0)=\varphi_{rel}(0,0,\infty)=\infty$. We assumed that the $R$, $v_{rot}$ and $v_{mac}$ contributions to $\varphi_{rel}$ were separable, and we find that this is approximately true. 

We first considered the $R$ dependence of $\varphi_{rel}$ using a Kurucz92-based spectrum of a a 5750K, $\log(g)=4.5$, [Fe/H]=0.0, Sun-like star split up into 100\,\AA\ spectral chunks. We used a Kurucz92-based spectrum -- instead of a BT-Settl spectrum -- because of the extremely fine wavelength spacing available with the Kurucz92 spectra. Our Kurucz92 spectrum had a wavelength spacing of 0.001\,\AA, as compared to a median spacing of 0.05\,\AA\ in the BT-Settl spectrum. Though the BT-Settl spacing gives a well sampled spectrum for most applications, we will see that for our specific examination of line broadening mechanisms the 0.05\,\AA\ spacing has a noticeable effect, by effectively setting a base spectral resolution of $R$$\approx$105,000 (see Figure \ref{fig:fitabs}).   

\begin{figure}[t]
\vskip -0.0in 
\epsscale{1.2} 
\plotone{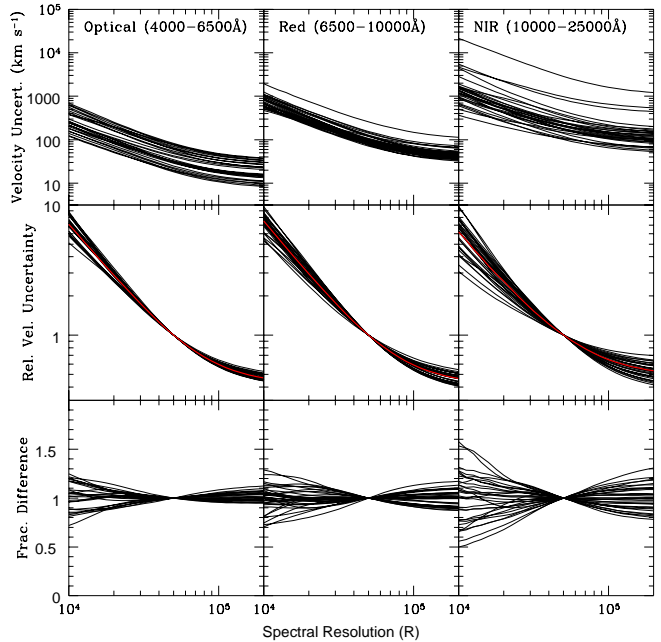}
\vskip -0.0in 
\caption{The velocity uncertainty of 100\,\AA\ chunks as a function of spectral resolution using Kurucz92-based spectra. The top row shows the absolute uncertainty for the chunks in our three wavelength regions, while the middle row shows the velocity uncertainty for each chunk normalized to $R$=50,000. The red line in the middle panel is the median of the relative velocity uncertainties. The bottom row shows the fractional difference between each individual chunk's relative uncertainties and this median. The middle row demonstrates that all of the wavelength chunks respond roughly self-similarly to changes in spectral resolution.}
\label{fig:res}
\end{figure}

We assumed that the effect of instrumental spectral resolution could be approximated by convolving a spectrum with a Gaussian of FWHM equal to $c/R$. We therefore convolved each 100\,\AA\ chunk with Gaussians corresponding to a range of $R$-values, and calculated the velocity uncertainties. 

To do this, we first chose a continuum SNR per velocity bin of unity, by setting $N_i=1$ in the continuum (and with corresponding lower values of $N_i$ in the absorbed portions of the spectrum). We then applied a given spectral resolution ($\varphi_{rel}[R,0,0]$) via Gaussian convolution, calculated the slope at each bin in the synthetic spectrum, $(dN/dV)|_i$, and used Equation (\ref{eq:2040}) to calculate the velocity uncertainty for the chunk as a function of $R$.

Figure \ref{fig:res} shows the absolute values of the chunk velocity uncertainties for a 5750K star from $R$=10,000 to $R$=200,000 in the top row. We have separated the chunks into ``optical'', ``red'' and ``NIR''. In absolute terms there is a range in uncertainties across the chunks as a result of specific spectral features in specific locations.

To check for any wavelength dependencies, we normalized the curves from the top row of Figure \ref{fig:res} by their value at $R$=50,000. The middle row of Figure \ref{fig:res} shows these normalized curves, and the red line shows their median value at each spectral resolution. One effect to note is that the dispersion of the chunks in the middle row panels increases as one moves to the red; that is, while the optical chunks all behave very similarly, the NIR chunks show more variation relative to each other. Specifically, many of the chunks in the NIR seem to be less affected by changing resolution than the optical chunks.

This occurs because the less affected chunks, which are mostly at longer wavelengths, have residual molecular features in the model spectra we are using. Since the lines that make up the molecular bandheads are very closely spaced, they blend together into composite lines with a large widths even at high resolutions. This lessens the effect of increasing spectral resolution in resolving these features.

The red line in the middle row of Figure \ref{fig:res} is the median relative velocity uncertainty across all the chunks, and the bottom row of Figure \ref{fig:res} shows the fractional difference between all the chunks and this median. The fractional difference across the entire optical and red wavelength ranges is rarely more than 20\%, while the NIR chunks stay within about 30\% of the calculated chunk median.

We repeated the above procedure to numerically calculate the median velocity uncertainty vs. $R$ using all the Kurucz92-derived spectra from 4000K to 7500K in steps of 250K. Across this temperature range the results for the optical, red, and NIR chunks were similar to our illustrative, 5750K, example. The velocity uncertainties of the chunks roughly stayed within 25\% of the calculated median.

\begin{figure}[t]
\vskip -0.0in 
\epsscale{1.2} 
\plotone{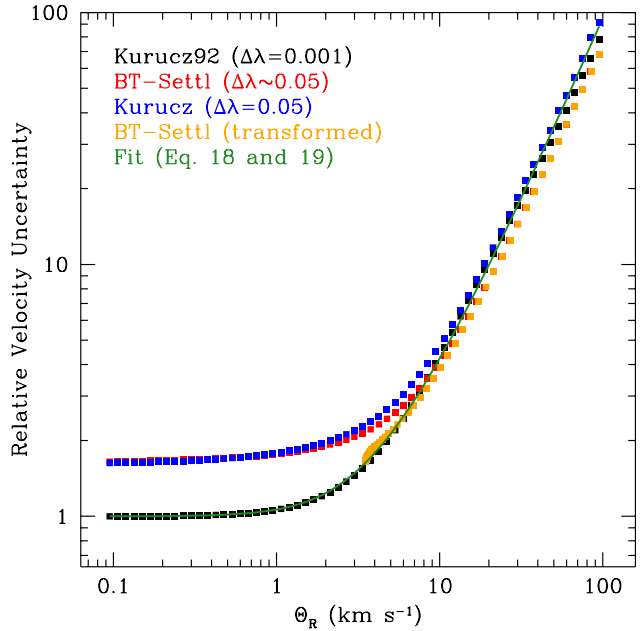}
\vskip -0.0in 
\caption{Velocity uncertainty, relative to the velocity uncertainty in a Kurucz92-based spectra at $R$=$3\times10^6$, plotted as a function of $\Theta_R=c/R$. The black points show how the uncertainty on Kurucz92-based spectra with a wavelength spacing of $\Delta\lambda=0.001$ varies with resolution, and the green line shows our best fit using Equation (\ref{eq:3140}). In addition, note the difference between the black Kurucz92-based spectra and a similar analysis done on the BT-Settl-based spectra (red points). The wavelength spacing of the BT-Settl spectra ($\Delta\lambda\sim0.05$) imposes a resolution ``floor'' of $R$$\approx$105,000. This is demonstrated by the orange points, which show the behavior of the BT-Settl spectra if if we include a base of $R$=105,000 and add this in quadrature to the resolution increase being applied by our Gaussian convolution, thereby imposing a resolution floor of $\Theta_R=2.9$\,km s$^{-1}$. This transformed BT-Settl behaves similarly to the black Kurucz92 line. Alternatively, the blue points show how a Kurucz92 spectrum sampled with a $\Delta\lambda=0.05$ wavelength spacing, comparable to the BT-Settl spectra, reacts to changing spectral resolution.}
\label{fig:fitabs}
\end{figure}

To determine the $R$ dependence of $\varphi_{rel}$, we fit to the chunk median velocity uncertainties. Since we are interested in the relative change in velocity uncertainty for the chunks, we re-normalized each chunk median so that the velocity uncertainty at $R$=$3\times10^6$ was unity, such that $\varphi_{rel}(\infty,0,0)=1$. To describe the average FWHM of the spectral lines in the chunks, we fit the chunk median as a Voigt profile with some inherent width $\Theta_0$, such that the relative increase in the average FWHM of the chunks scaled as $\varphi_{rel}^{3/2}$, with
\begin{equation}\label{eq:3140}
\varphi_{rel} = \frac{0.5346\Theta_0 + \sqrt{0.2166\Theta_0^2 + \Theta_{R}^2}}{\Theta_0},
\end{equation}
where $\Theta_{R}=c/R$. For each wavelength region, we fit the measured median velocity uncertainties as a function of resolution using Equations (\ref{eq:3110}) and (\ref{eq:3140}) and by finding the best-fit value of $\Theta_0$. Figure \ref{fig:fitabs} shows the median chunk velocity uncertainty in the optical for a 5750K star as a function of $\Theta_{R}$ in black, overlaid by with our best-fit in green. The residuals to the best-fit are no more than 3\% across the entire range of resolutions at 5750K.

\begin{figure}[t]
\vskip -0.0in 
\epsscale{1.2} 
\plotone{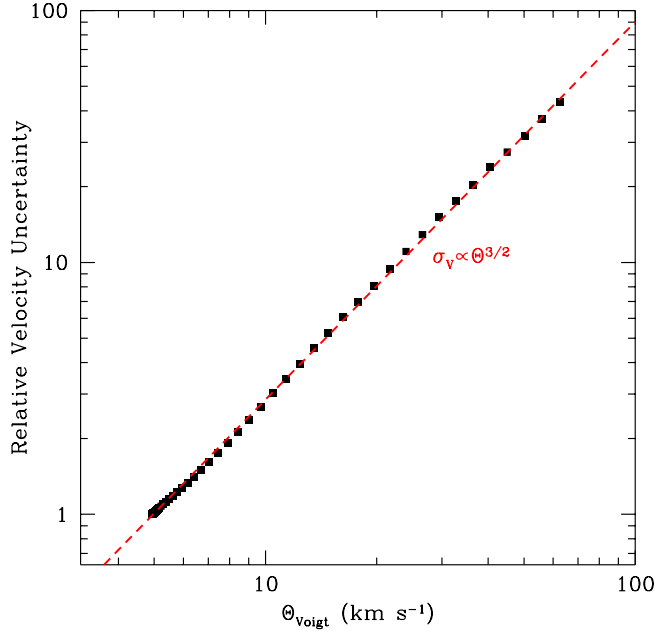}
\vskip -0.0in 
\caption{The median velocity uncertainty of a 5750K, $\log g=4.5$ star, relative to the velocity uncertainty for that same star at $R$=$1\times10^6$, as a function of the width of the underlying Voigt profile (Equation [\ref{eq:2180}]). This figure highlights that the velocity uncertainty is not set by the resolution alone, but by the combination of the underlying average line width and the spectral resolution. The Voigt width used here is the combination of the average inherent line width in the spectrum from Equation (\ref{eq:3150}) and a changing spectral resolution, from $R$=$1\times10^6$ to $R$=10,000. While the calculated points deviate from  $\sigma_V\propto\Theta_{Voigt}^{3/2}$ on a small scale, the overall best fit to the median relative velocities is given by a $\Theta_{Voigt}^{1.49}\approx\Theta_{Voigt}^{3/2}$ scaling.}  
\label{fig:logvoigt}
\end{figure}

As an illustration of the effects of the different wavelength spacing in the Kurucz92 spectra and the BT-Settl spectra, Figure \ref{fig:fitabs} also shows a similarly calculated curve for the velocity uncertainty of a 5800K, $\log(g)=4.5$ BT-Settl spectrum in red. Note that the BT-Settl curve asymptotes to a significantly higher velocity uncertainty at high resolution (low $\Theta_{R}$). We interpret this as a result of the coarser wavelength spacing in the BT-Settl spectrum, which imposes a base ``resolution'' of $R\approx5250\mathrm{\AA}/0.05\mathrm{\AA}=105,000$ in the optical. Indeed, if we include a base of $R$=105,000 and add this in quadrature to the resolution increase being applied by our Gaussian convolution, thereby imposing a resolution floor of $\Theta_R=2.9$\,km s$^{-1}$, the red BT-Settl curve in Figure \ref{fig:fitabs} transforms to the orange line in Figure \ref{fig:fitabs} and nearly matches the Kurucz92-based results. As a further test we also generated a Kurucz92 spectrum with a 0.05\,\AA\ wavelength spacing, and its curve is shown in blue. This coarser Kurucz92 spectrum nearly matches the BT-Settl results, which makes us confident that the difference between the 0.001\,\AA\ Kurucz92 spectrum and the 0.05\,\AA\ BT-Settl spectrum is primarily a result of the different wavelength spacings. Figure \ref{fig:fitabs} also illustrates why we used the Kurucz92-based spectra for our examination of line broadening mechanisms: the unbroadened BT-Settl spectra are not sampled finely enough to represent $\varphi_{rel}(\infty,0,0)=1$. 

We note that at low-$R$ in Figure \ref{fig:fitabs} there is an offset between the Kurucz92 and BT-Settl spectra. We were not able to completely determine the cause of this offset, which is equal to about 0.15 dex at $R$=3,000 and about 0.05 dex at $R$=30,000.   

\subsubsection{Comparison to Other Work}

It is also worth noting at this point that other authors' \citep[e.g.,][]{hatzes1992,bouchy2001,bottom2013} numerical calculations of the dependence of how velocity uncertainty scales with spectral resolution find that at low-$R$ the uncertainty goes approximately as $\sigma_V\propto R^{-1}$ \citep{hatzes1992,bouchy2001} or $\sigma_V\propto R^{-1.2}$ \citep{bottom2013}. There are two connected points to consider here. First, as we have seen, we mathematically expect the velocity uncertainty to scale as $\Theta^{1.5}$, where $\Theta$ is set by \emph{both} $\Theta_{R}$ and the inherent line width $\Theta_0$. This means that considering only $\Theta_{R}$, as these authors do, does not account for the effect of the intrinsic width of the lines on the velocity uncertainty.

Second, as we shall see, for a Sun-like star in the optical, $\Theta_{R}$ dominates the inherent line width (i.e., $\Theta_{R}\gtrsim10\Theta_0$) only for resolutions less then 6,000. Thus \cite{bouchy2001} and \cite{bottom2013}, who consider down to $R$=10,000, find $\sigma_V\propto R^{-1}$ and $\sigma_V\propto R^{-1.2}$, respectively, since they are largely fitting over the transition regime between $\Theta_{R}$ and $\Theta_0$. \cite{hatzes1992} directly measure the velocity uncertainty at $R$=2,500 and find $\sigma_V\propto R^{-1}$, but we believe this result to be a poor fit to their measurements, since this line passes substantially underneath the $R$=2,500 point. Indeed, if we take the three points in Figure 1 of \cite{hatzes1992} and fit them using our formalism, we recover $\sigma_V\propto R^{-1.5}$ at low resolution.  

\begin{figure}[t]
\vskip -0.0in 
\epsscale{1.2} 
\plotone{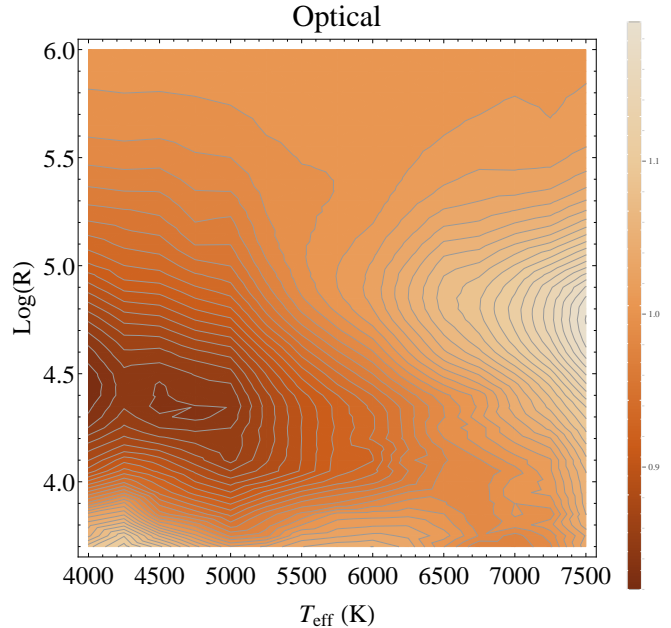}
\vskip -0.0in 
\caption{The ratio of our fits for the velocity uncertainties, calculated using Equations (\ref{eq:3140}) and (\ref{eq:3150}), to the numerically computed chunk uncertainties as a function of changing spectral resolution and effective temperature in the optical. The RMS across all resolutions and temperatures is 7\%, and the peak value is about 20\%.}  
\label{fig:resopt}
\end{figure}

If we plot the median relative velocity uncertainty for a 5750K, $\log g=4.5$, star against the underlying Voigt width (see Equation [\ref{eq:2180}]) instead of the width solely due to spectral resolution (Figure \ref{fig:logvoigt}), we can immediately see that the best fit to the calculated uncertainties goes as $\sigma_V\propto\Theta^{3/2}$. 

\subsubsection{Temperature Dependence of $\Theta_0$}

We repeated our procedure of fitting the velocity uncertainty as a function of $\Theta_R$ for the full temperature range covered by the Kurucz92 models (4000K to 7500K), and for the three wavelength ranges we consider (optical, red, and NIR). We found that the best-fit value of $\Theta_0$ decreased roughly linearly with temperature and was slightly different in each regime. Specifically, $\Theta_0$ goes as 
\begin{eqnarray}\label{eq:3150}
\Theta_0 &=& 5.10521\,\mathrm{km\ s}^{-1}\ (1-0.6395\,\Delta T_{eff})\\
&&\ \ \ \ \ \ \ \ \ \ \ \ \ \ \ \ \ \ \ \ \ \ \ \ \ \ \ \ \ \ \ \ \ \ \ \mathrm{for\ 4000\ to\ 6500\,\AA} \nonumber \\ 
\Theta_0 &=& 3.73956\,\mathrm{km\ s}^{-1}\ (1-0.1449\,\Delta T_{eff}) \nonumber \\
&&\ \ \ \ \ \ \ \ \ \ \ \ \ \ \ \ \ \ \ \ \ \ \ \ \ \ \ \ \ \ \ \ \ \,\mathrm{for\ 6500\ to\ 10000\,\AA} \nonumber \\
\Theta_0 &=& 6.42622\,\mathrm{km\ s}^{-1}\ (1-0.2737\,\Delta T_{eff}) \nonumber \\
&&\ \ \ \ \ \ \ \ \ \ \ \ \ \ \ \ \ \ \ \ \ \ \ \ \ \ \ \ \ \ \ \ \mathrm{for\ 10000\ to\ 25000\,\AA}, \nonumber
\end{eqnarray}
where $\Delta T_{eff}=T_{eff}/5800\mathrm{K}-1$. These equations are good to 3\% in the optical and 5\% in the red and NIR.

\begin{figure}[t]
\vskip -0.0in 
\epsscale{1.2} 
\plotone{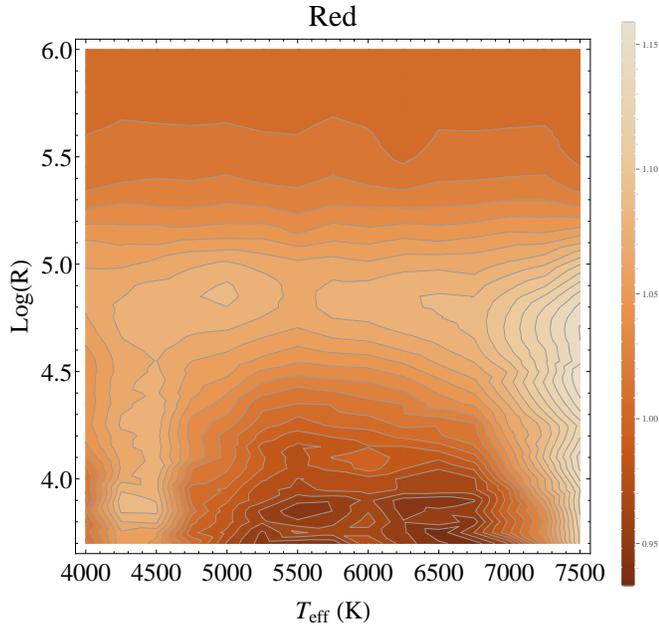}
\vskip -0.0in 
\caption{The ratio of our fits for the velocity uncertainties, calculated using Equations (\ref{eq:3140}) and (\ref{eq:3150}), to the numerically computed chunk uncertainties as a function of changing spectral resolution and effective temperature in the optical. The RMS across all resolutions and temperatures is 4\%, and the peak value is about 10\%.}  
\label{fig:resred}
\end{figure}

\begin{figure}[t]
\vskip -0.0in 
\epsscale{1.2} 
\plotone{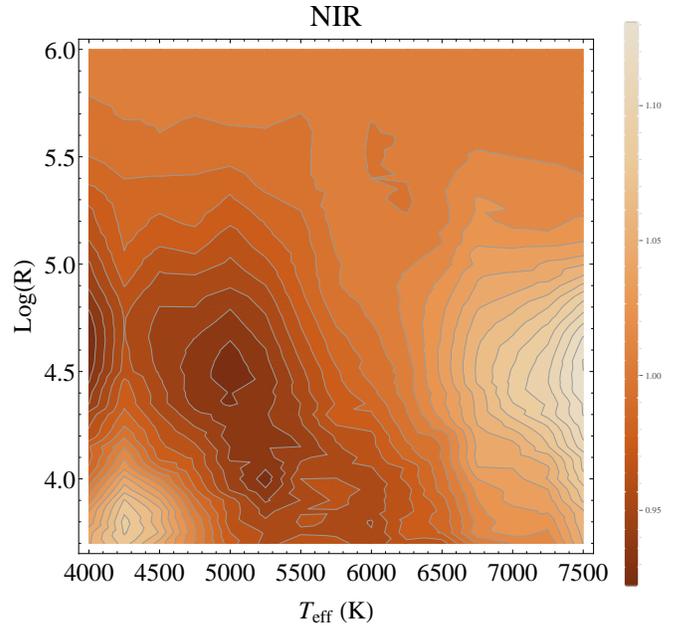}
\vskip -0.0in 
\caption{The ratio of our fits for the velocity uncertainties, calculated using Equations (\ref{eq:3140}) and (\ref{eq:3150}), to the numerically computed chunk uncertainties as a function of changing spectral resolution and effective temperature in the optical. The RMS across all resolutions and temperatures is 3.5\%, and the peak value is about 10\%.}  
\label{fig:resnir}
\end{figure}

Figures \ref{fig:resopt}, \ref{fig:resred}, and \ref{fig:resnir}, show the ratio of our fits
to the numerically calculated chunk velocity uncertainties as a function of effective temperature and spectral resolution. Across the entire range of temperature and resolution the ratios have a standard deviation of about 5\% about unity. The largest differences occur in the optical at very low and very high temperatures for resolutions near 60,000. Our fits under-predict the uncertainties at low temperatures and over-predict at high temperatures, both by about 20\% at these specific locations.

\subsubsection{Stellar Rotation}

We next turned to the effect of stellar rotation on the velocity uncertainty. We presumed that the effect of changing $v_{rot}$ is similar to spectral resolution $R$, in that convolving a spectrum with a rotation kernel is similar to convolving a spectrum with a Gaussian, and that we could describe the change in the line width similarly as
\begin{equation}\label{eq:3155}
\varphi_{rel} = \frac{0.5346\Theta_0 + \sqrt{0.2166\Theta_0^2 + \Theta_{rot}^2}}{\Theta_0}.
\end{equation}
Here $\Theta_{rot}$ is the FWHM of the rotation kernel. Equation (\ref{eq:3155}) presumes a Gaussian FWHM measurement, so for the purposes of Equation (\ref{eq:3155}) we will need to calculate an ``equivalent Gaussian width'' for the rotation kernel. This is the width of a Gaussian which, when convolved with a spectrum, will cause the equivalent broadening and associated increase in the velocity uncertainty as the rotational kernel. As we have seen in Section 2.1, this substitution may be easily made: the increase in velocity uncertainty caused by Gaussian broadening differs from rotational broadening only by a leading numerical coefficient.

To determine the equivalent Gaussian width of the rotation kernel, we set Equations (\ref{eq:2160}) and (\ref{eq:3192}) equal to each other and solve for $\Theta_G$. Thus,
\begin{equation}\label{eq:3194}
\Theta_{G,eq} \approx \left(\frac{0.347+0.146\,\epsilon}{0.69}\right)^{2/3}\,\Theta_{rot}.
\end{equation}
As we noted in Section 2.1, the velocity uncertainty caused by the rotational broadening at a particular rotation velocity is only weakly dependent on the precise value of $\epsilon$. The difference in the proportionality constant relating $\Theta_{rot}^{3/2}$ and $\sigma_V$ varies by about 10\% from minimum to maximum. We will therefore take the average value, which occurs at $\epsilon=0.5$, for all of our results. This makes $\Theta_{G,eq} = 0.72\,\Theta_{rot}$.

\begin{figure}[b]
\vskip -0.0in 
\epsscale{1.2} 
\plotone{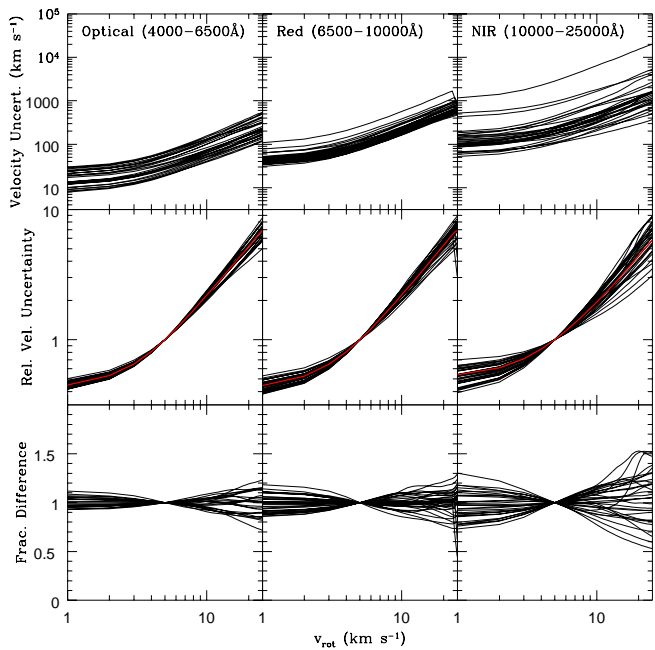}
\vskip -0.0in 
\caption{The velocity uncertainty of 100\,\AA\ chunks as a function of stellar rotation velocity from Kurucz92-based spectra. The top row shows the absolute uncertainty for the chunks in our three wavelength regions, while the middle row shows the velocity uncertainty for each chunk normalized to $v_{rot}$=5 km s$^{-1}$. The red line in the middle panel is the median of the relative velocity uncertainties. The bottom row shows the fractional difference between each individual chunk's relative uncertainties and this median. The middle row demonstrates that all of the wavelength chunks respond roughly self-similarly to changes in stellar rotation.}
\label{fig:vrot}
\end{figure}

Having determined $\Theta_{G,eq}$ for rotation, we know wish to know how velocity uncertainty scales with stellar rotation velocity. In a manner similar to how we approached spectral resolution, we calculated how the velocity uncertainty in the wavelength chunks changed as $v_{rot}$ went from 0 to 25 km/s using Kurucz92-based spectra. We used the {\sc avsini} routine packaged with the {\sc spectrum} code to apply the rotation kernel to our spectra using with $\epsilon=0.5$. As one can see in Figure \ref{fig:vrot}, changing $v_{rot}$ is similar to changing spectral resolution in that it largely effects all of the wavelength chunks in the same way. The middle row of Figure \ref{fig:vrot} shows the relative change in the velocity uncertainty normalized to $v_{rot}=5$ km s$^{-1}$ for a 5750K, $\log(g)=4.5$, [Fe/H]=0.0 star in our three wavelength regimes. Similar to our approach to fitting the effect of changing spectral resolution, we also calculated median values for our entire temperature range.

We then fit the chunk averages in the same manner as for spectral resolution. In doing so, we found that rotational velocity affects the relative velocity uncertainty of a spectrum in almost exactly the same way as does spectral resolution. That is, when $\Theta_{G,eq}=\Theta_{R}$ the velocity uncertainty is almost exactly the same across our entire temperature range.

\subsubsection{Macroturbulence}

In addition to rotation, we also considered the effect of macroturbulence in the stellar atmosphere. For simplicity we assumed simple isotropic macroturbulence, so that the effect of macroturbulence with velocity $v_{mac}$ is the same as convolving a spectrum with a normalized Gaussian with standard deviation $v_{mac}/2$. Under this assumption the FWHM of the macroturbulence kernel is then simply 
\begin{equation}\label{eq:3170}
\Theta_{mac} = 2\sqrt{2\ln 2}\ \frac{v_{mac}}{2} \approx 1.18\,v_{mac}.
\end{equation}

Note that in reality, the effects of rotation and macroturbulence are difficult to observationally separate when $v_{rot}\approx v_{mac}$ \citep[e.g.,][]{valenti2005}. This is partially a result of the fact that the effect of real macroturbulence is not isotropic, and partially because macroturbulent and rotational broadening are observed as a disk-integrated broadening profile. This makes the two effects difficult to separate observationally.  

Typical macroturbulent velocities for field dwarfs are on the order of a few km s$^{-1}$ \citep{valenti2005,gray2008,bruntt2010}, with mid F-dwarfs at about 6 km s$^{-1}$ and decreasing linearly with spectral type to about 1.5 km s$^{-1}$ for an early K-dwarf. We used the empirical relation for $v_{mac}$ as a function of temperature determined by \cite{bruntt2010}, which we list along with other stellar properties in Section 4.

We therefore will use the same results we had for spectral resolution (Voigt line profiles, temperature dependence) and apply it to rotation and macroturbulence. Putting this all together, we may rewrite Equation (\ref{eq:3110}) as
\begin{eqnarray}\label{eq:3180}
\varphi_{rel} &=& \Biggr(\frac{0.5346\Theta_0(T_{eff})}{\Theta_0(T_{eff})} \\
&+& \frac{\sqrt{0.2166\Theta_0^2 + \Theta_{R}^2 + 0.518\,\Theta_{rot}^2 + \Theta_{mac}^2}}{\Theta_0(T_{eff})}\Biggr)^{3/2}. \nonumber
\end{eqnarray}

\subsection{Temperature: BT-Settl Based}

Temperature affects both the width and the number of lines usable for radial velocity measurements in a spectrum, and unlike the line broadening mechanisms considered in above, stellar temperature should be considered an intrinsic part of line generation. Without a detailed treatment of how spectral lines are created, it is therefore difficult to arrive at a physically motivated analytic expression for how the velocity uncertainty in a spectrum changes along with effective temperature. While the thermal velocity width of the lines will scale simply as the square-root of the effective temperature, the pressure of the atmospheric layer where these lines are generated will change as well. These two competing effects -- temperature width and pressure width -- are not easily separable. In addition, the number of lines in a spectrum depends upon a host of factors such as opacities, atomic energy levels, and ionization equilibria that also provide no simple scaling with temperature.

\begin{figure}[b]
\vskip -0.0in 
\epsscale{1.2} 
\plotone{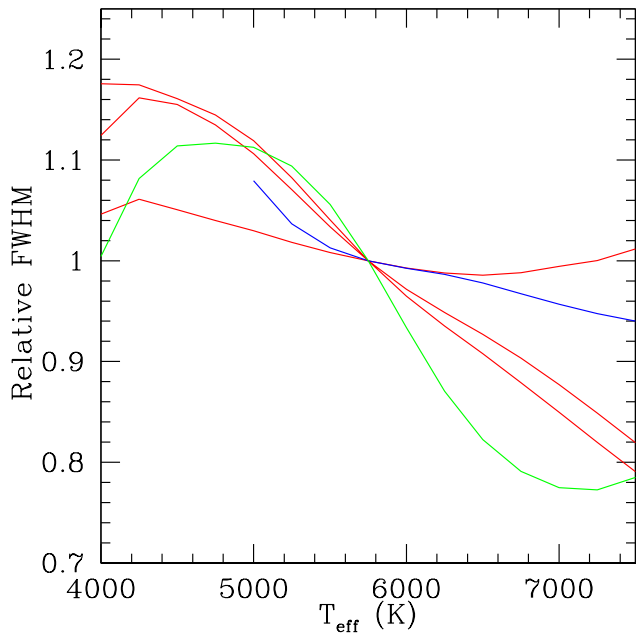}
\vskip -0.0in 
\caption{Measured FWHMs of specific isolated lines as a function of effective temperature. We have normalized each to be unity at 5800K, so as to identify any broad trends. The lines are color-coded according to the atom responsible: red are iron lines at 5294.5\,\AA, 5905.7\,\AA, and 6078.5\,\AA, blue is a silicon line at 6125.0\,\AA, and green is a nickel line at 6482.7\,\AA. While the FWHMs of all the lines generally grow smaller as temperature increases, one can see the large variation in how specific lines react. This is in contrast to, for example, spectral resolution, which will alter the FWHM of lines uniformly.}
\label{fig:teffline}
\end{figure}

Figure \ref{fig:teffline} is an illustration of the complexity of the issue. We selected three isolated iron lines, one silicon line, and one nickel line, and measured the FWHM of these lines as a function of temperature in a series of [Fe/H]=0, $\log(g)=4.5$, Kurucz92 spectra. Our naive expectation was that the FWHM of all the lines would behave self-similarly and increase as the square-root of the effective temperature. Instead, the measured FWHMs decreased with effective temperature and display an idiosyncratic temperature dependence. As mentioned above, we attribute this divergence from our expected behavior to a set of competing effects, including varying local pressure, differences in ionization levels, and differences in opacities.

We therefore determined a purely numerical and descriptive scaling for how the relative velocity uncertainty in a spectrum changes with effective temperature. Due to their availability over a greater range of temperatures we used the BT-Settl models for this fitting. We again used 100\,\AA\ chunks sliced out of spectra with effective temperatures of 2600K to 7600K and $\log g=4.5$ in the three wavelength ranges we are considering.   

\begin{figure}[t]
\vskip -0.0in 
\epsscale{1.2} 
\plotone{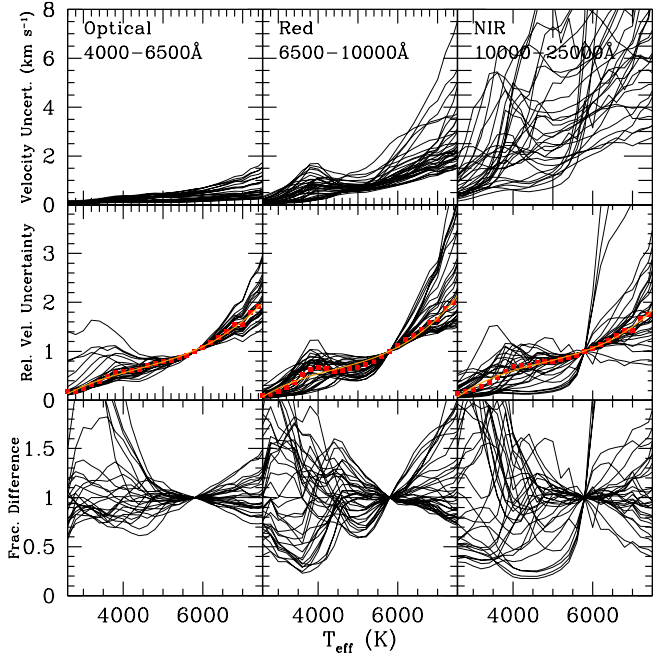}
\vskip -0.0in 
\caption{The velocity uncertainty of 100\,\AA\ chunks as a function of stellar effective temperature using BT-Settl-based spectra. The top row shows the absolute uncertainty for the chunks in our three wavelength regions, while the middle row shows the velocity uncertainty for each chunk normalized that for 5800K. The red points in the middle panel is the median of the relative velocity uncertainties, while the orange overplotted line is our fit to the median. The bottom row shows the fractional difference between each individual chunk's relative uncertainties and this median. The chunks show considerably more relative differences as a function of temperature compared to changing spectral resolution or rotation. This reflects the complicated competing effects that changing temperature causes in spectral line generation, including changing pressure levels, ionization states, and opacities.}
\label{fig:teff}
\end{figure}

The top panels of Figure \ref{fig:teff} show the effect of changing temperature on the velocity uncertainty for each of the chunks. It is immediately apparent from Figure \ref{fig:teff} that changing the effective temperature acts in a much less self-similar way across the chunks as compared to the external line-broadening mechanisms we considered previously. Not surprisingly, while the behavior of the optical chunks is roughly self-similar (left side of the middle row of Figure \ref{fig:teff}), the NIR chunks (right side of the middle row of Figure \ref{fig:teff}) show considerable differences. This is largely caused by the different line generation mechanisms at optical and NIR wavelengths. While the optical is mostly populated by atomic lines that change strength relatively slowly with effective temperature, the NIR chunks possess more molecular lines that have a sharp temperature dependence. For example, while a 4000K, $\log(g)=4.5$, spectra from 24500\,\AA\ to 24600\,\AA\ is a forest of CO molecular lines, that same 100\,\AA\ chunk in a 7000K, $\log(g)=4.5$, spectrum has only one atomic Fe and one atomic Mg line as its major spectral features.

This also illustrates the vital importance of choosing the appropriate wavelength range in the NIR when designing an RV survey. For example, our illustrative 24500\,\AA\ to 24600\,\AA\ chunk is a perfect example of a wavelength region that would be a reasonable choice for an RV survey focusing on K and M stars, but it would be a poor choice for a NIR survey that would observe FGK dwarfs. We consider the choice of wavelength range in more detail in the discussion section.      

\begin{figure}[b]
\vskip -0.0in 
\epsscale{1.2} 
\plotone{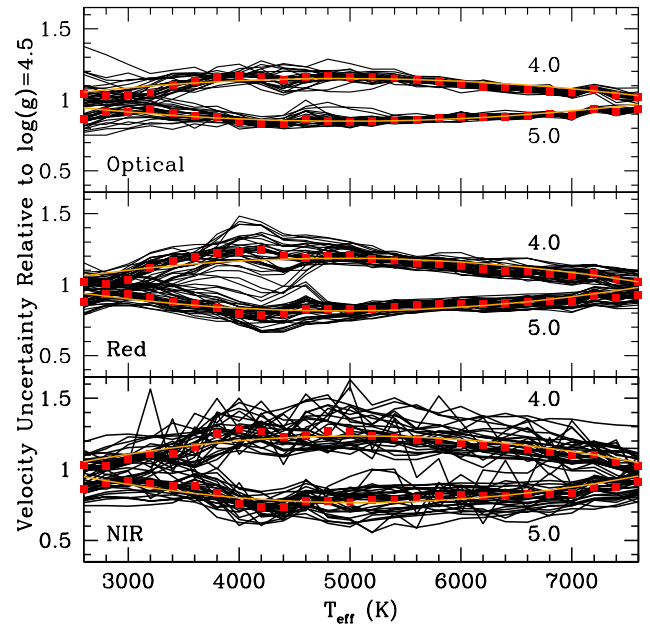}
\vskip -0.0in
\caption{Velocity uncertainty as a function of effective temperature and surface gravity, relative to $\log(g)=4.5$ using BT-Settl spectra. The black lines are individual 100\,\AA\ spectral chunks, and the red points are the medians of all the chunks at the effective temperatures of the model spectra. The overplotted orange lines are our fits to the chunk medians. Notice that at the extremes of the temperate range we consider, where we expect the stellar surface gravity to be most different from $\log(g)=4.5$, the effect of changing surface gravity is the least. Surface gravity therefore plays a minor ($\sim$10\%) in setting velocity uncertainties.}
\label{fig:logg}
\end{figure}

To generally describe the behavior of the wavelength chunks as a function of temperature, we roughly approximated the chunk medians by taking a least squares polynomial fit to the three wavelength regions, which yielded
\begin{eqnarray}\label{eq:3210}
f(T_{eff})_{\mathrm{Opt}} &=& 1+2.04515\,\Delta T_{eff}+3.13362\,\Delta T_{eff}^2\\
&&\ \ \ \ \ \ \ \ \ \ \ \ \ \ \ \ \ \ \ \ \ \ \ \ \ \ \ \ \ \ \ \ \ +4.23845\,\Delta T_{eff}^3 \nonumber \\
f(T_{eff})_{\mathrm{Red}} &=& 1+2.18311\,\Delta T_{eff}+4.00361\,\Delta T_{eff}^2 \nonumber \\
&&\ \ \ \ \ \ \ \ \ \ \ \ \ \ \ \ \ \ \ \ \ \ \ \ \ \ \ \ \ \ \ \ \ +5.62077\,\Delta T_{eff}^3 \nonumber \\
f(T_{eff})_{\mathrm{NIR}} &=& 1+1.62418\,\Delta T_{eff}+2.62018\,\Delta T_{eff}^2 \nonumber \\
&&\ \ \ \ \ \ \ \ \ \ \ \ \ \ \ \ \ \ \ \ \ \ \ \ \ \ \ \ \ \ \ \ \ +5.01776\,\Delta T_{eff}^3, \nonumber
\end{eqnarray}
where $\Delta T_{eff}=T_{eff}/5800\mathrm{K}-1$. The bottom panels of Figure \ref{fig:teff} show the fractional difference between each chunk's relative velocity uncertainties and the chunk medians. The relations in Equation (\ref{eq:3210}) replicate the chunk medians to with 4\%. Relative to the medians, while the optical chunks are relatively coherent, one can see the moving towards redder wavelengths causes the chunks to vary more about the median, for the reasons outlined above. 

\subsection{Surface Gravity: BT-Settl Based}

Surface gravity, through pressure broadening effects, can change both the depth and width of spectral lines, and thus the RV velocity uncertainty of a spectrum. The exact response of lines to changes in surface gravity is dependent upon several factors. For example, a decrease in gravity can cause a weak line to either gain or lose strength depending upon the ionization state of the atoms. We therefore approached surface gravity effects in a manner similar to effective temperature, by determining a purely numerical and descriptive scaling of velocity uncertainty.

\begin{figure}[b]
\vskip -0.0in 
\epsscale{1.2} 
\plotone{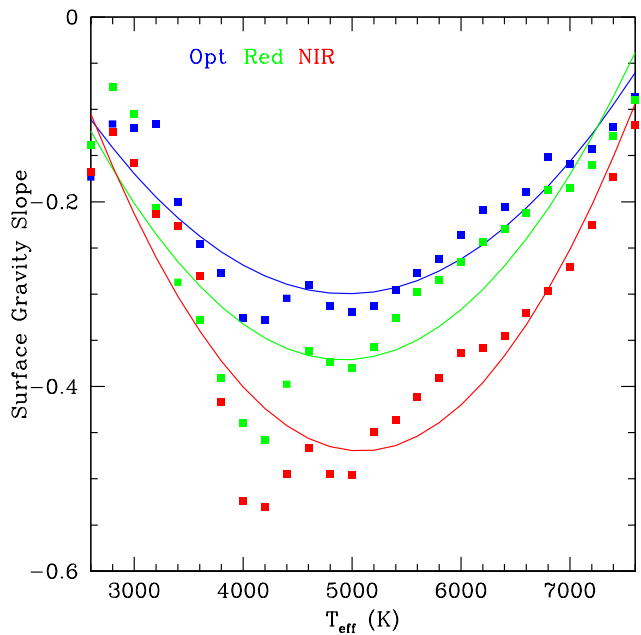}
\vskip -0.0in
\caption{The points show the measured median slopes used in Equation (\ref{eq:3310}) to describe the change in velocity precision as a linear function of the change in surface gravity ($\log g$). The overplotted lines are the fits to these measurements, as per Equation (\ref{eq:3320}).}
\label{fig:loggslope}
\end{figure}

We took BT-Settl spectra from 2600K to 7600K with surface gravities of $\log(g)=4.0$ and $\log(g)=5.0$ and calculated the velocity uncertainties of each relative to $\log(g)=4.5$. (Figure \ref{fig:logg}). We again split each spectrum up into 100\,\AA\ chunks and examined all three of our wavelength bands. Figure \ref{fig:logg} displays the results for the individual chunks, as well the median velocity uncertainty across all the chunks, and our fit to the median as a function of effective temperature and surface gravity. All three bands behaved roughly the same, with the greatest relative difference occurring around 4500K and the smallest differences happening towards the ends of the temperature range.

Since the chunk medians in all three bands are nearly symmetric about unity, we decided to describe the change in velocity error as a linear function of surface gravity relative to what the velocity uncertainty would be for $\log g=4.5$. Specifically,
\begin{equation}\label{eq:3310}
f(\log g) = m\cdot\Delta \log(g) + 1
\end{equation}
with $\Delta \log(g) = \log g - 4.5$. The slope $m$ depends upon the effective temperature and the band observed,
\begin{eqnarray}\label{eq:3320}
m_{\mathrm{Opt}} &=& -0.27505\,(1-1.22211\,\Delta T_{eff}\\
&&\ \ \ \ \ \ \ \ \ \ \ \ \ \ \ \ \ \ \ \ \ \ \ -4.17622\,\Delta T_{eff}^2) \nonumber \\
m_{\mathrm{Red}} &=& -0.33507\,(1-1.41362\,\Delta T_{eff} \nonumber \\
&&\ \ \ \ \ \ \ \ \ \ \ \ \ \ \ \ \ \ \ \ \ \ \ -4.63727\,\Delta T_{eff}^2) \nonumber \\
m_{\mathrm{NIR}} &=& -0.43926\,(1-1.12505\,\Delta T_{eff} \nonumber \\
&&\ \ \ \ \ \ \ \ \ \ \ \ \ \ \ \ \ \ \ \ \ \ \ -4.53938\,\Delta T_{eff}^2), \nonumber
\end{eqnarray}
where $\Delta T_{eff}=T_{eff}/5800\mathrm{K}-1$. Figure \ref{fig:loggslope} shows the measured values of $m$ across the temperature range, overplotted by the fits from Equation (\ref{eq:3320}). Note that these results are only accurate for surface gravities between $4.0\le\log g\le5.0$.

Since the extremes of our considered temperature range, where we expect $|\Delta \log(g)|$ to be the largest, show the smallest effect due to changing gravity, we expect surface gravity to play a relatively minor role in determining the velocity error of a spectrum. Indeed, when we include how we expect surface gravity to vary with stellar mass (Equation [\ref{eq:4050}]) we find that changing surface gravity is never more than a 10\% effect, and is frequently less.

\subsection{Metallicity: Kurucz92 Based}

Finally, we consider the effect of differing metallicity, which we parameterized as [Fe/H], on velocity uncertainties. Similarly to the effect of changing effective temperature and changing surface gravity, we expected that changing metallicity would alter the velocity uncertainty in a spectrum in a way that is difficult to capture analytically from \emph{a priori} arguments. We therefore confined ourselves to a numerical, descriptive, scaling for the effect of changing [Fe/H].

To do so, we used three sets of Kurucz92-model spectra at fixed effective temperatures of 5000K, 5750K, and 6500K. All three sets had a fixed surface gravity of $\log g=4.5$, no $\alpha$-enhancement, and metallicities of [Fe/H]=[-2.0,-1.5,-1.0,-0.5,+0.0,+0.5]. We used Kurucz92 based spectra, instead of BT-Settl spectra, due to the availability of Kurucz92 models with a wide range of [Fe/H] values. The BT-Settl spectra, while more physically motivated in terms of what metallicities are available for a given effective temperature and surface gravity, do not provide an arbitrary range of metallicities with no $\alpha$-enhancement. As a result, our results are confined to the optical wavelengths, from 4000\,\AA\ to 6500\,\AA, where the Kurucz92 line-lists are robust.

\begin{figure}[t]
\vskip -0.0in 
\epsscale{1.2} 
\plotone{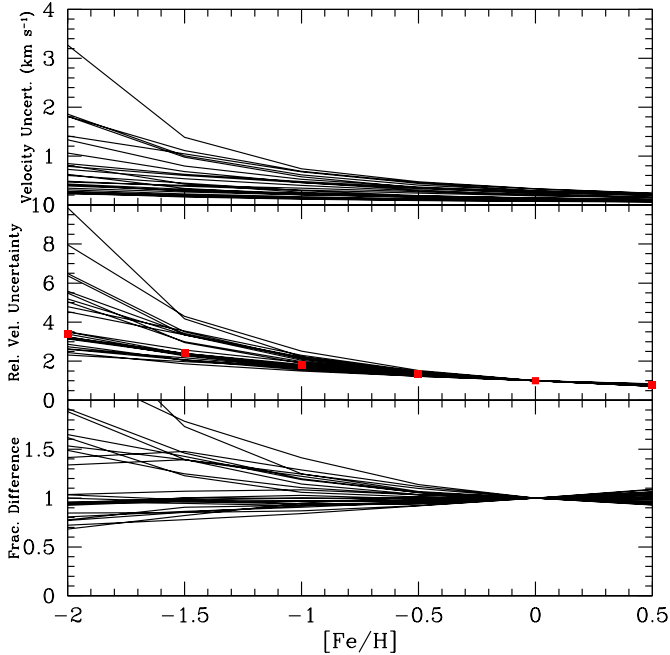}
\vskip -0.0in
\caption{The velocity uncertainty of 100\,\AA\ chunks as a function of metallicity using Kurucz92-based spectra. These results are for a star with a fixed effective temperature of 5750K and a fixed $\log g=4.5$ in the optical (4000\,\AA\ to 6500\,\AA). The top panels shows the absolute uncertainty for the chunks, while the middle row shows the velocity uncertainty for each chunk normalized to [Fe/H]=0.0. The red points in the middle panel are the median of the relative velocity uncertainties. The bottom row shows the fractional difference between each individual chunk's relative uncertainties and this median. As expected, a decrease in [Fe/H] reduces the number of lines and average line strength in a spectrum, thus increasing the velocity uncertainty.}
\label{fig:feh}
\end{figure}

Figure \ref{fig:feh} shows the results of varying the metallicity on the velocity uncertainties for a 5750K, $\log g=4.5$ star in the optical. The results for the 5000K and 6500K spectra were similar, differing by at most 15\% at the low metallicity end. As one would expect, the velocity uncertainties for all of the chunks increases as [Fe/H] decreases; a result of the absorption features in the spectra becoming weaker and less numerous. Some of the specific 100\,\AA\ chunks are strongly effected by this, as their major lines are drastically diminished at [Fe/H]=-2.0. 

We fit to the red chunk median in the middle panel of Figure \ref{fig:feh} to find the change in velocity uncertainty relative to [Fe/H]=0.0. This gave
\begin{equation}\label{eq:30410}
f(\mathrm{[Fe/H]}) = 10^{-0.27\,\mathrm{[Fe/H]}}.
\end{equation}
For all three effective temperatures, this result is accurate to 15\% over the range of [Fe/H] values we considered, with the highest difference occurring at [Fe/H]=-2.0.  

\subsection{Final Expressions}

Putting together all of our results from the preceding analysis, we arrive at a semi-analytic expression for the velocity uncertainty using an arbitrary number of 100\,\AA\ chunks as a function of $R$, $v\sin i$, $T_{eff}$, $\log g$, and [Fe/H]:
\begin{eqnarray}\label{eq:3410}
\sigma_{V} &=&  \frac{1}{\sqrt{\sum\limits \frac{I_{0,i}}{\sigma_{V,i}^2}}} \Biggr(\frac{0.5346\Theta_0(T_{eff})}{\Theta_0(T_{eff})} \\
&+& \frac{\sqrt{0.2166\Theta_0^2 + \Theta_{R}^2 + 0.518\,\Theta_{rot}^2 + \Theta_{mac}^2}}{\Theta_0(T_{eff})}\Biggr)^{3/2} \nonumber \\
&\times& f(T_{eff})\ f(\log g)\ f(\mathrm{[Fe/H]}). \nonumber
\end{eqnarray}

\begin{deluxetable*}{c|ccccccccccc}
\tablecaption{Chunk Velocity Uncertainties (km/s)}
\tablewidth{6in}
\startdata
\hline
        & 2600K  & 2800K  & 3000K  & 3200K  & 3400K  & 3600K  & 3800K  & 4000K  & 4200K  & ...\\
\hline
4000\,\AA & 0.0681 & 0.0756 & 0.0821 & 0.0919 & 0.0930 & 0.0856 & 0.0911 & 0.0821 & 0.0758 & ...\\
4100\,\AA & 0.0783 & 0.0829 & 0.0848 & 0.0916 & 0.0873 & 0.0821 & 0.0739 & 0.0655 & 0.0556 & ...\\ 
4200\,\AA & 0.1232 & 0.1175 & 0.1236 & 0.1413 & 0.1387 & 0.1302 & 0.1186 & 0.1059 & 0.0887 & ...\\
4300\,\AA & 0.0487 & 0.0597 & 0.0769 & 0.0930 & 0.1062 & 0.1106 & 0.1109 & 0.0999 & 0.0923 & ...\\
...     & ...    & ...    & ...    & ...    & ...    & ...    & ...    & ...    & ...    & ...
\enddata
\tablecomments{The complete table is available as online data at http://www.personal.psu.edu/tgb15/beattygaudi/table1.dat.}
\end{deluxetable*}

Where $\Theta_0(T_{eff})$ is given in Equation (\ref{eq:3150}), $f(T_{eff})$ in Equation (\ref{eq:3210}), $f(\log g)$ in Equation (\ref{eq:3310}) and $f(\mathrm{[Fe/H]})$ in Equation (\ref{eq:30410}). The leading summation term is a sum over all of the 100\,\AA\ chunks observed, with $\sigma_{V,i}$ as the velocity uncertainty of the individual chunks for $R$=$\infty$, $v_{rot}$=0, $v_{mac}$=0, and for a continuum level of 1 photon per velocity element. The true continuum level in each chunk is incorporated via $\sqrt{I_{0,i}}$. Note that we are defining the continuum here in velocity-space, and not pixel-space as is conventional in the observational literature. 

\begin{figure}[t]
\vskip -0.0in 
\epsscale{1.2} 
\plotone{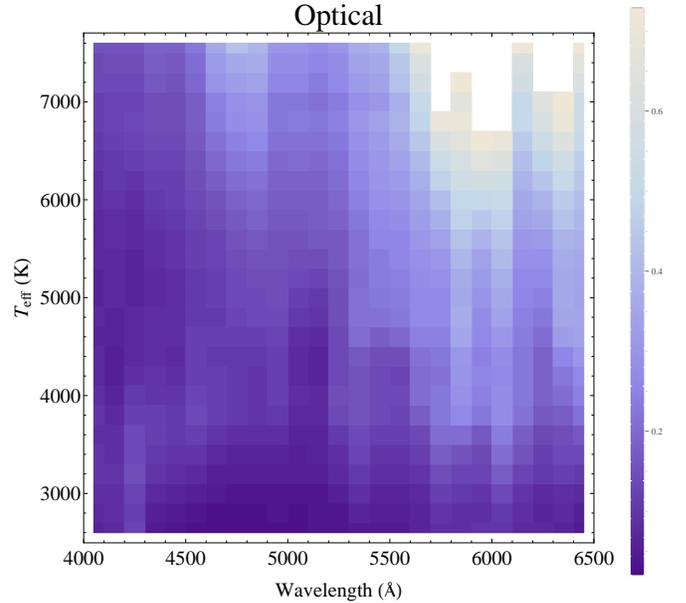}
\vskip -0.0in
\caption{The radial velocity uncertainty (in km s$^{-1}$) for individual 100\,\AA\ chunks in the optical, over the temperature range we consider. This assumes a continuum flux in each chunk equal to unity, or in terms of Equation (\ref{eq:3410}), that $I_{0,i}=1$ photons per km s$^{-1}$. This is a display of the information in Table 1. The darker regions indicate a lower velocity uncertainty at a given wavelength and effective temperature.}
\label{fig:optq}
\end{figure}

\begin{figure}[t]
\vskip -0.0in 
\epsscale{1.2} 
\plotone{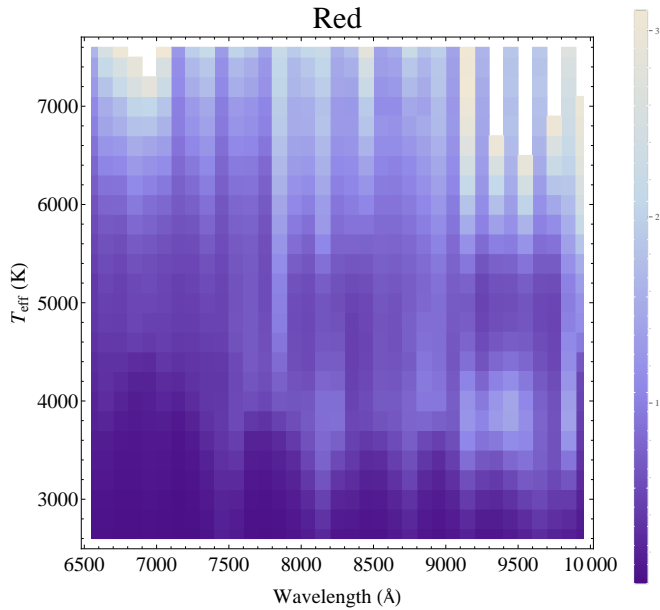}
\vskip -0.0in
\caption{The radial velocity uncertainty (in km s$^{-1}$) for individual 100\,\AA\ chunks in the red, over the temperature range we consider. This assumes a continuum flux in each chunk equal to unity, or in terms of Equation (\ref{eq:3410}), that $I_{0,i}=1$ photons per km s$^{-1}$. This is a display of the information in Table 1. The darker regions indicate a lower velocity uncertainty at a given wavelength and effective temperature.}
\label{fig:redq}
\end{figure}

\begin{figure}[t]
\vskip -0.0in 
\epsscale{1.2} 
\plotone{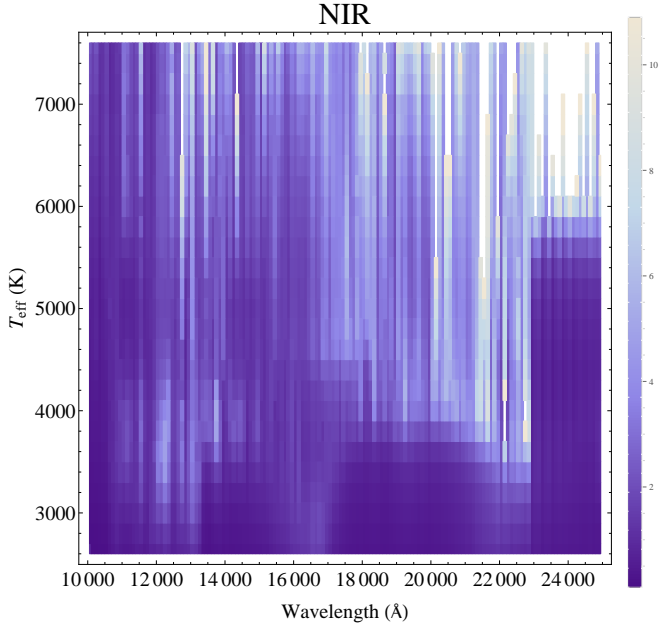}
\vskip -0.0in
\caption{The radial velocity uncertainty (in km s$^{-1}$) for individual 100\,\AA\ chunks in the NIR, over the temperature range we consider. This assumes a continuum flux in each chunk equal to unity, or in terms of Equation (\ref{eq:3410}), that $I_{0,i}=1$ photons per km s$^{-1}$. This is a display of the information in Table 1. The darker regions indicate a lower velocity uncertainty at a given wavelength and effective temperature. The sharp feature around 23000\,\AA\ is the CO molecular bandhead.}
\label{fig:nirq}
\end{figure}

Table 1 lists values of $\sigma_{V,i}$, in km s$^{-1}$, normalized to $I_{0,i}=1$ photon per velocity element, for 100\,\AA\ chunks between 4000\,\AA\ and 25000\,\AA\ for temperatures between 2600K and 7600K and a constant $\log g$=4.5. This information is also displayed graphically for our three wavelength regions, in Figures \ref{fig:optq}, \ref{fig:redq}, and \ref{fig:nirq}. Table 1 was calculated using the BT-Settl spectra, which used the \cite{asplund2009} solar abundances and had solar metallicity with no $\alpha$-enhancement. The values in Table 1 are normalized for a continuum level of 1 photon per km s$^{-1}$. Spectroscopic observations usually quote their SNR per pixel ($SNR_{pix,i}$), which can be converted into the appropriate units for $I_{0,i}$ by taking
\begin{equation}\label{eq:3430}
I_{0,i} = \frac{(SNR_{pix,i})^2\ n_{pix,i}}{\Delta V_{chunk,i}},
\end{equation} 
where $\Delta V_{chunk,i}$ is the velocity span of the wavelength chunk, and $n_{pix,i}$ is the number of pixels on the detector used to observe the chunk. 

\section{Stellar Properties}

Having so far considered the dependence of velocity uncertainties on stellar parameters independently of what is physically reasonable, we know wish to apply this formalism towards real stars. Our focus here is to determine what is the dominant source of the velocity uncertainty for main-sequence stars.  

We have chosen to focus on main sequence stars so that we may use stellar mass as a single variable to then calculate all of the stellar properties that determine velocity uncertainty. As above, the stellar properties we are interested in are rotation velocity, macroturbulence, effective temperature, and surface gravity. Additionally, we also wish to know the overall bolometric luminosities and radii of the stars. The first allows us to calculate the continuum level of the spectra, while the second will be necessary to estimate rotation velocities from the rotation periods predicted from stellar gyrochronology relations. To that end, we fit relations for effective temperature, luminosity, radius, and surface gravity from the measurements listed in Table 1 of \cite{torres2010} for stars cooler than 7600K:
\begin{equation}\label{eq:4020}
T_{eff} = 5603\mathrm{K} \left(\frac{M_*}{M_\odot}\right)^{0.41} \approx 5800\mathrm{K} \left(\frac{M_*}{M_\odot}\right)^{0.5},
\end{equation}
\begin{equation}\label{eq:4040}
L_* = 1.06 L_\odot \left(\frac{M_*}{M_\odot}\right)^{4.48} \approx 1.0 L_\odot \left(\frac{M_*}{M_\odot}\right)^{4.5},
\end{equation}
\begin{equation}\label{eq:4030}
R_* = 1.12 R_\odot \left(\frac{M_*}{M_\odot}\right)^{1.12},
\end{equation}
and
\begin{equation}\label{eq:4050}
\log(g) = 4.96 - 0.58\left(\frac{M_*}{M_\odot}\right) \approx 5 - 0.5\left(\frac{M_*}{M_\odot}\right).
\end{equation}
For all the functions except the radius relation we have also given rough approximations, which we use when simplifying our numeric results. We note that these relations are only roughly consistent with each other, a result of our collapsing stars of different main sequence ages and metallicities onto single relations. Unlike temperature, luminosity, radius, and surface gravity, the rotation velocity does not display a simple scaling with stellar mass, and so we consider it separately and in more detail in the next section.

For the macroturbulent velocity, we used the empirical relation for $v_{mac}$ as function of effective temperature determined by \cite{bruntt2010},
\begin{equation}\label{eq:4060}
v_{mac} = 1.976\,\mathrm{km\ s}^{-1} + 16.14\,\Delta T_{eff} + 19.713\,\Delta T_{eff}^2,
\end{equation}
where we have substituted $\Delta T_{eff}=T_{eff}/5800\mathrm{K}-1$. \cite{bruntt2010} make the point that this relation is only valid for stars with $\log(g)>4.0$ and between 5000K to 6500K. We use Equation (\ref{eq:4060}) to estimate the macroturbulent velocity for stars from 5000K to 7600K, and set a constant macroturbulent velocity of 0.51\,km $s^{-1}$ (the value of Equation (\ref{eq:4060}) at 5000K) for all stars cooler than 5000K. Above 6500K, thus should not introduce large errors into our results, because the rotational velocities of these hot stars are at least five times larger than the calculated macroturbulent velocities. 

In addition to these stellar properties, there is also RV ``jitter'' in stars, which causes additional uncertainty in precision velocity measurements. Astrophysical jitter is generally a result of either star-spots on the stellar photosphere, or short-period solar-like asteroseismic oscillations. For the latter, many of the existing RV surveys mitigated the effect of asteroseismic jitter by integrating on a star for longer than the oscillation periods, which are typically about 5 minutes. Unfortunately, the jitter caused by star-spots has no comparable solution, other than avoiding stars with high activity indices. Typical jitter values for main-sequence stars cooler than 6300K and with average activity levels are measured to be around 3 to 4 m s$^{-1}$ \citep{wright2005,isaacson2010,martinez2010}. As has been noted by all of these authors, this undoubtedly includes jitter from astrophysical and instrumental sources. Indeed, \cite{isaacson2010} find that their measured jitter in K-dwarfs is completely uncorrelated with stellar activity; they therefore conclude that the jitter displayed by these stars is likely a result of instrumental effects.

Since velocity jitter is caused by additional astrophysical and instrumental sources, the appropriate way to incorporate it into our formalism is to add the jitter value for a given star in quadrature to the Poisson velocity uncertainty calculated using Equation (\ref{eq:3410}). RV jitter is therefore not a component we need to consider in determining the dominant sources of Poisson velocity uncertainty in main-sequence stars. We thus leave it aside for now, other than to note the importance of jitter in using our results to fully model a realistic RV survey.

\subsection{Stellar rotation}

Since stellar rotation strongly broadens stellar lines, we undertook a detailed examination of the true rotational speeds, $v_{\mathrm{rot}}$, of stars within our mass range. In general, stars with substantial outer convective envelopes, from $0.4M_\odot$ to the Kraft Break \citep{kraft1970} at $1.1M_\odot$ will magnetically brake over the first billion years of their lives and coalesce onto a single mass-rotation-age relation. This is the basis of stellar gyrochronology \citep{barnes2003}. Stars less massive than $0.4M_\odot$ generally do not brake effectively, and so do not evolve onto a single mass-rotation-age relationship. Similarly, stars more massive than the Kraft Break mass of $1.1 M_\odot$ have very thin outer convective envelope and retain almost all of their primordial angular momentum. These stars will slightly lengthen their rotational periods due to the gradual increase of their radii on the main sequence, but this is change is on the order of 2\% over their lifetime. These heavier stars also do not, therefore, evolve onto a single mass-rotation-age relationship.
 
For stars between $0.4M_\odot$ and $1.1 M_\odot$ we used the modified Kawaler spin down model developed by \cite{epstein2014} to determine the rotation periods of stars at a certain mass and age. We then used the mass-radius relation in Equation (\ref{eq:4030}) to convert the rotation periods into rotation speeds. For stars older than 0.5 Gyr this spin-down model predicts a tight mass-rotation-age relation down to $0.4M_\odot$, with more scatter as one goes to younger ages and lower mass. We linearly interpolated between the available model grid points in mass and age and took the median rotation period as the rotational period of all the stars with that mass and age.

For stars less massive than $0.4M_\odot$, we treated $v_{rot}$ as a distribution, with velocities uniformly distributed in velocity between zero and some upper bound $v_{\mathrm{max}}$. This roughly replicates the distribution of M-dwarf rotation velocities observed by \cite{reiners2012}. The upper bound was set equal to the Kawaler rotation velocity at the high mass end, $v_{rot}(0.4M_\odot)$, which is an age dependent quantity, and increased linearly with mass through 10 km s$^{-1}$ at $0.2M_\odot$. Thus
\begin{eqnarray}\label{eq:4107}
v_{\mathrm{rot}}(M_*) &=& \biggr([10\,\mathrm{km\ s}^{-1}-v_{rot}(0.4M_\odot)] \\
&&\times \frac{0.4M_\odot-M_*}{0.2M_\odot}\biggr)+v_{rot}(0.4M_\odot) \nonumber \\
&&\ \ \ \ \ \ \ \ \ \ \ \ \ \ \ \ \ \ \ \ \mathrm{for}\ M_*<0.4M_\odot. \nonumber
\end{eqnarray}

We treated stars more massive than $1.1 M_\odot$ in a similar manner. Based on the observations and discussion in \cite{gaige1993} and \cite{reiners2003}, we treated the $v\sin i$ distribution of stars heavier than $1.1M_\odot$ as uniformly distributed in velocity between zero and a mass dependent upper bound. For the massive stars, this upper bound was set to the Kawaler rotation velocity at 1.1$M_\odot$, $v_{rot}(1.1M_\odot)$, and the bound increased linearly with mass through 100 km s$^{-1}$ at $1.5M_\odot$. Therefore for massive stars we have
\begin{eqnarray}\label{eq:4108}
v_{\mathrm{max}}(M_*) &=& \biggr([100\,\mathrm{km\ s}^{-1}-v_{rot}(1.1M_\odot)] \\
&& \times \frac{M_*-1.1M_\odot}{0.4M_\odot}\biggr)+v_{rot}(1.1M_\odot) \nonumber \\
&&\ \ \ \ \ \ \ \ \ \ \ \ \ \ \ \ \ \ \ \ \mathrm{for}\ M_*>1.1M_\odot. \nonumber
\end{eqnarray}
 
\section{Dominant Sources of Velocity Uncertainty}

We now wish to assess the dominant astrophysical sources of velocity uncertainty in RV measurements of main sequence field stars. To do so, we calculated the change in the velocity uncertainty caused by the $I_{0,i}$, $\Theta$, $T_{eff}$, $\log g$, and [Fe/H] terms in Equation (\ref{eq:3410}). This gives us the relative change in velocity uncertainty as a function of stellar properties, independent of the specific wavelength chunk (or chunks) chosen. The specific wavelength information is provided in  Equation (\ref{eq:3410}) by the $\sigma_{V}$ values given in Table 1, and serves to simply set the appropriate absolute value of the uncertainty.  

We begin by calculating how the relative velocity uncertainty scales as a function of stellar mass in the wavelength ranges we consider. We used the relations in Section 4 to determine the effective temperature, surface gravity, and luminosity as a function of mass. For the luminosity, we included the effect of overall changes in the bolometric luminosity, normalized to a 5800K star, by Equation (\ref{eq:4040}), and the effect on the observed luminosity caused by the shifting of the blackbody emission across the specific wavelength range being observed. We refer to this as the ``blackbody effect.'' This blackbody term means that the exact results will still depend on the specific wavelength chunk used to calculate the relative uncertainties. For each of our three bands (optical, red, and NIR), we used a 100\,\AA\ chunk in the middle of the wavelength range to calculate the blackbody effect. In our tests, using chunks at the extreme of our wavelength bands changes the calculated uncertainties by 5\% or less, and does not affect our ultimate conclusions.         

As a fiducial example, we set $R$=60,000, [Fe/H]=0.0, and the stellar age to 2.0$\,$Gyr. Recall that the age will set the rotation velocity of stars between 0.4$\,M_\odot$ and 1.1$\,M_\odot$, with younger stars rotating more rapidly. We chose 2.0$\,$Gyr so as to be broadly representative of a typical field FGK dwarf in the Solar neighborhood \citep{nordstrom2004}. 

\begin{figure}[t]
\vskip -0.0in 
\epsscale{1.2} 
\plotone{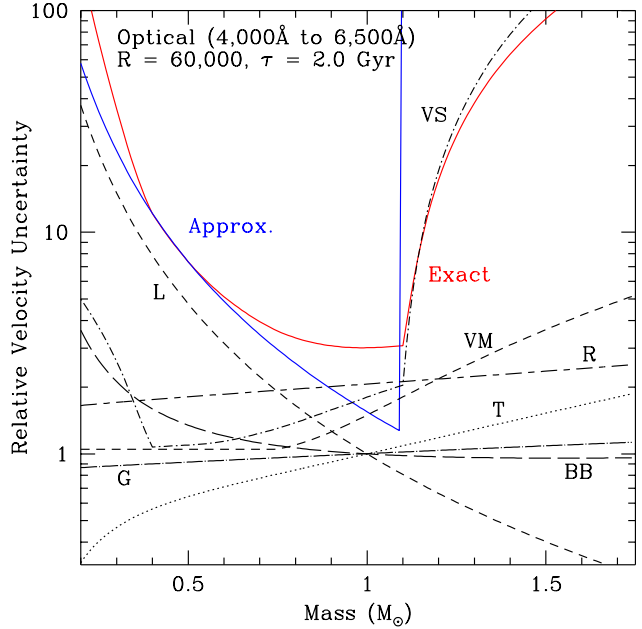}
\vskip -0.0in
\caption{The velocity uncertainty of observations of ``realistic'' stars, using the stellar properties described in Section 4, as a function of stellar mass in the optical. The uncertainties are shown relative to the uncertainty of an observation of a ``perfect'' non-rotating Sun-like star with no macroturbulence and observed using $R$$\sim$$\infty$. The red line shows the exact uncertainty calculation for observations using a spectral resolution of $R$=60,000 and a stellar age of $\tau$=2.0 Gyr. We have decomposed the overall velocity uncertainty into its constituent parts (black lines), as described in the third paragraph of Section 5. These are: changes to line strengths and numbers due to effective temperature (T), changes in the overall bolometric luminosity (L), luminosity changes from the blackbody peak shifting relative to the spectral bandpass (BB), stellar rotation (VS), macroturbulence (VM), surface gravity changes (G), and spectral resolution ($R$). The blue approximation line is from Equation (\ref{eq:4210}), and has been normalized to match the calculated uncertainty at 0.5$M_\odot$. In the optical, the approximated uncertainty is roughly proportional to the overall luminosity (L).}
\label{fig:erropt}
\end{figure}

\begin{figure}[t]
\vskip -0.0in 
\epsscale{1.2} 
\plotone{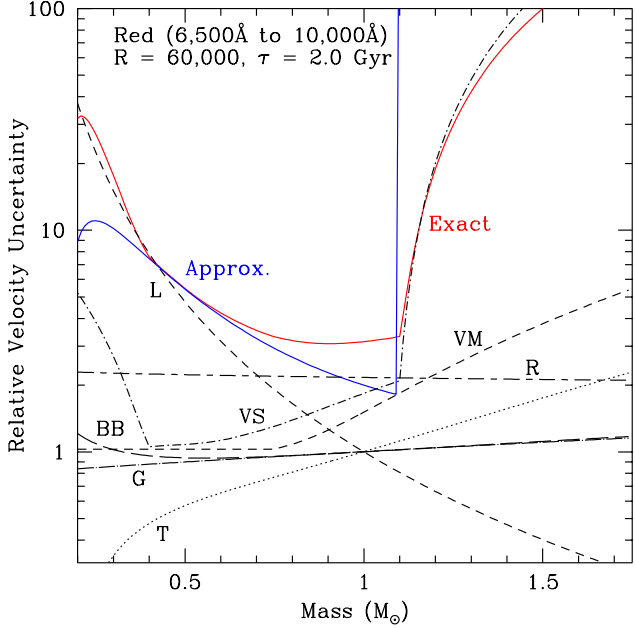}
\vskip -0.0in
\caption{The velocity uncertainty of observations of ``realistic'' stars, using the stellar properties described in Section 4, as a function of stellar mass in the red. The uncertainties are shown relative to the uncertainty of an observation of a ``perfect'' non-rotating Sun-like star with no macroturbulence and observed using $R$$\sim$$\infty$. The red line shows the exact uncertainty calculation for observations using a spectral resolution of $R$=60,000 and a stellar age of $\tau$=2.0 Gyr. We have decomposed the overall velocity uncertainty into its constituent parts (black lines), as described in the third paragraph of Section 5 and in the caption of Figure 19. The blue approximation line is from Equation (\ref{eq:4210}), and has been normalized to match the calculated uncertainty at 0.5$M_\odot$. In the red, the approximated uncertainty is roughly proportional to the overall luminosity (L) and temperature effects (T).}
\label{fig:errred}
\end{figure}

\begin{figure}[t]
\vskip -0.0in 
\epsscale{1.2} 
\plotone{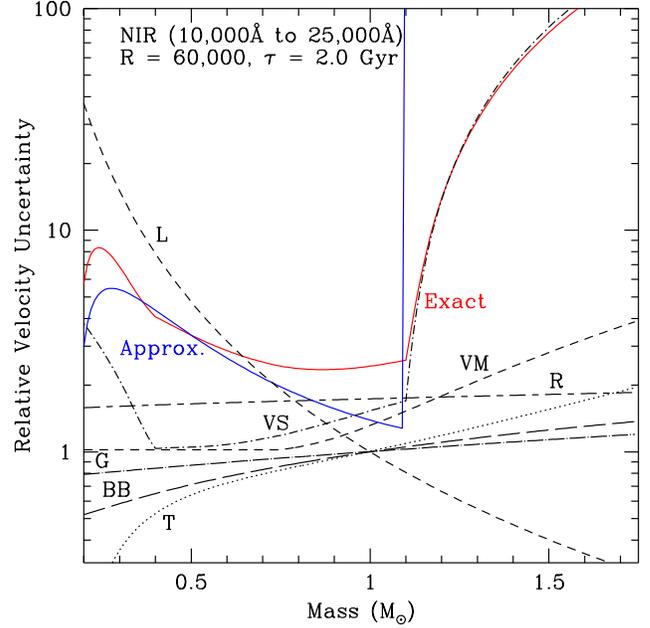}
\vskip -0.0in
\caption{The velocity uncertainty of observations of ``realistic'' stars, using the stellar properties described in Section 4, as a function of stellar mass in the NIR. The uncertainties are shown relative to the uncertainty of an observation of a ``perfect'' non-rotating Sun-like star with no macroturbulence and observed using $R$$\sim$$\infty$. The red line shows the exact uncertainty calculation for observations using a spectral resolution of $R$=60,000 and a stellar age of $\tau$=2.0 Gyr. We have decomposed this overall velocity uncertainty into its constituent parts (black lines), as described in the third paragraph of Section 5 and in the caption of Figure 19. The blue approximation line is from Equation (\ref{eq:4210}), and has been normalized to match the calculated uncertainty at 0.5$M_\odot$. In the NIR, the approximated uncertainty is roughly proportional to the overall luminosity (L), temperature effects (T), and the blackbody effect (BB).}
\label{fig:errnir}
\end{figure}

Figures \ref{fig:erropt}, \ref{fig:errred}, and \ref{fig:errnir} show the expected uncertainty as a function of mass for the optical, red, and NIR as the solid red line. To illustrate how the overall uncertainty is determined by the underlying stellar parameters, these figures also show how the uncertainty changes if we fix all but one of the physical processes that affect the velocity uncertainty and depend on stellar mass. These parameters are changes to line strengths and numbers due to effective temperature (T), changes in the overall bolometric luminosity (L), luminosity changes from the blackbody peak shifting relative to the spectral bandpass (BB), stellar rotation (VS), macroturbulence (VM), surface gravity changes (G), and spectral resolution ($R$). 

There are two things to immediately note. First, below $0.4M_\odot$ and above $1.1M_\odot$ the VS curve is for the maximum observed $v\sin i$ at each mass. Second, the changing effect of spectral resolution as a function of mass is a result of the average inherent line widths varying with effective temperature, per Equation (\ref{eq:3150}). Larger inherent line widths (e.g., at lower temperatures in the optical) cause finite spectral resolution to have a smaller effect on the velocity uncertainties.

In general, these three figures demonstrate that there are two general regimes for the RV errors of F-M main sequence stars: luminosity and temperature dominated uncertainties below $1.1M_\odot$ when stellar rotation is low, and rotation dominated uncertainties for stars above that mass. In particular, the rapid increase in the average rotation for more massive stars means that in all three bands the maximum velocity uncertainty raises sharply in this regime, becoming an order of magnitude larger than it would be for a Sun-like star at just $\approx1.25M_\odot$.  

Below $1.1M_\odot$, on the other hand, the three wavelength regions behave differently. This primarily due to the changing effect of the blackbody peak shifting relative to the observed bandpass (the BB line in all three figures). In the optical (Figure \ref{fig:erropt}) the blackbody effect causes larger uncertainties for lower mass stars as the peak shifts into the red, particularly below $0.6M_\odot$. In the red (Figure \ref{fig:errred}), the blackbody term is nearly constant, as the peak is moving through this wavelength regime. By the time we reach the NIR (Figure \ref{fig:errnir}), the blackbody term finally begins to reduce the velocity uncertainties of lower mass stars relative to solar-mass stars.

To illustrate this different behavior in the different bandpasses, and to understand the dominant source of velocity uncertainty in a conceptually straightforward manner, we derived simple approximations to how the velocity uncertainty of low-mass ($M_*<1.1M_\odot$) stars scaled with stellar mass. In the optical, the blackbody effect nearly cancels the effect of changing effective temperature, leaving the bolometric luminosity as the dominant source of uncertainty. In the red, where the blackbody effect is nearly constant, we approximate the velocity uncertainty as a combination of luminosity and temperature effects. Finally, in the NIR, we must also include the blackbody effect into our approximation using luminosity and temperature. We find that in the NIR the BB term is well fit by a simple linear relation with stellar mass, such that $\sigma_{V,\mathrm{BB}}\propto 0.4(M_*/M_\odot)+0.6$.

For our approximations for the relative uncertainty scaling below $1.1M_\odot$ we therefore have:
\begin{eqnarray}\label{eq:4210}
\sigma_{V,\mathrm{Opt}} &\propto& \left(\frac{L_*}{L_\odot}\right)^{-1/2}\ \ \ \ \ \ \ \ \ \ \ \ \ \ \ \: \mathrm{for}\ M_*<1.1M_\odot \\
\sigma_{V,\mathrm{Red}} &\propto& f_{Red}(T_{eff}) \left(\frac{L_*}{L_\odot}\right)^{-1/2}\ \mathrm{for}\ M_*<1.1M_\odot \nonumber \\
\sigma_{V,\mathrm{NIR}} &\propto& [0.4(M_*/M_\odot)+0.6]\; f_{NIR}(T_{eff}) \left(\frac{L_*}{L_\odot}\right)^{-1/2} \nonumber \\
&&\ \ \ \ \ \ \ \ \ \ \ \ \ \ \ \ \ \ \ \ \ \ \ \ \ \ \ \ \ \ \; \mathrm{for}\ M_*<1.1M_\odot. \nonumber
\end{eqnarray}
Above $1.1M_\odot$, where stellar rotation dominates, we would approximate the velocity error as $\sigma_{V}\propto\infty$, as the rapidly increasing rotation velocities widen the lines to unusability for detecting all but the most massive companions. Note that, in detail, since values of $v\sin i$ are observed to be uniformly distributed in this high-mass regime, some stars will have a $v\sin i$ low enough that rotation will not dominate the velocity uncertainties.  

The red lines in Figures 15, 16, and 17 show these approximations. Note that we have normalized each to match the directly calculated velocity uncertainty at 0.5$M_\odot$. This serves to account for the added uncertainty arising from our assumed value for spectral resolution, $R$, and the non-zero rotation of the low-mass stars. One can see the approximate velocity uncertainty agrees well with the calculated velocity uncertainty for lower mass stars, but the two diverge as mass increases above $M_*\approx0.8M_\odot$. This is a result of our ignoring the effects of stellar rotation in making our approximations.  

\section{Discussion}

In addition to the above approximations for the scaling of velocity uncertainty as a function of stellar mass, and the general scaling of velocity uncertainty with stellar parameters given in Section 3, there are three general points regarding RV surveys that are interesting to consider.

First, velocity precision does not scale linearly with $v \sin i$ for high rotation, or as $R$$^{-1}$ for low resolution, as has been claimed in some of the literature \citep[e.g.,][]{hatzes1992,connes1996,bouchy2001,bottom2013}. Instead, as we have shown, it goes appropriately as $(v \sin i)^{3/2}$, or $R^{-3/2}$. This arises due to two competing effects.  First, the intrinsic velocity information in the line scales with the FWHM $\Theta$ of the line as $\Theta^2$. Second, for weak lines, the photon noise is dominated by the continuum emission. The continuum spanned by the line is $\propto \Theta$, and thus the photon noise scales a $\Theta^{-1/2}$. The net result of these two effects results in the velocity uncertainty scaling as $\Theta^{3/2}$. This is in contrast to an isolated line with negligible continuum, where the precision scales simply as $\Theta$. Additionally, it is also important to remember that the velocity uncertainty is set by the overall line width, which is a combination of the inherent line width and the effects from broadening mechanisms like $v \sin i$ and $R$. By considering the total resulting line width, we are thus able to fit for and describe the velocity precision as a function of $v \sin i$ and $R$ over the entire range of these two parameters, and not just at the high $v \sin i$ or low $R$ limits.

Second, Figure \ref{fig:res} illustrates how arbitrarily increasing the spectral resolution if an instrument does not give arbitrarily low velocity uncertainties. Specifically, there is a ``knee'' in the uncertainty curves around $R$=60,000, after which increased spectral resolution has a much diminished effect. This occurs because once a spectrograph reaches $R$=60,000, the instrument is able to resolve almost all the lines in a stellar spectrum; further increases to $R$ therefore do not provide substantially more information. The precise spectral resolution we estimate for the position of the ``knee'' is similar to that determined by \cite{bouchy2001},but higher than the point of diminishing returns estimated in \cite{bottom2013}, who give a position of $R$=45,000. This difference is a result of the relatively coarse wavelength sampling in the BT-Settl spectra used by \cite{bottom2013}, which imposes an effective resolution floor of $R$$\approx$100,000, as we discuss in more detail in Section 3 and around Figure 5.

For stellar RV surveys, Figures \ref{fig:erropt}, \ref{fig:errred}, and \ref{fig:errnir} also show that there is a limited utility to spectral resolutions above $R$=60,000. One can see in these three figures that the effect of having $R$=60,000 (the horizontal dashed-line labeled ``R'') is the dominant source of velocity uncertainty for an extremely small range of masses. 

Instead, the dominant error source in radial velocity measurements will either be caused by luminosity ($M_*\lesssim0.8M_\odot$) or rotational velocity ($0.8M_\odot\lesssim M_*$). This means that relatively small detector arrays can be used effectively for multi-object RV surveys: 1024 pixels along the spectral dispersion axis would allow for a 100\,\AA\ spectral chunk to be imaged at $R$=60,000. As an example, if we ignore systematic and instrumental uncertainties, if such a survey observed a Sun-like star from 5100-5200\,\AA\ with SNR=200 per pixel, we predict that the Poisson velocity uncertainty on an individual observation would be about 6 m s$^{-1}$. 

More generally, we note that in terms of statistical velocity uncertainty there are sharply diminishing returns to be made from arbitrarily increasing spectral resolution. First, above a certain resolution all of the major lines in a spectrum will be resolved; for resolutions beyond this value the dominant source of uncertainty will be set by the line widths themselves. From Equation (\ref{eq:3150}), we can see that this will happen for a Sun-like star for $R$$\approx$60,000 in the optical, and $R$$\approx$80,000 and $R$$\approx$45,000 in the red and NIR, respectively. Second, stellar rotation will limit the utility of increased spectral resolution -- even in slowly rotating field stars. For $R$=60,000, for example, the effective line broadening caused by limited spectral resolution is equal to the amount of line broadening caused by 4.25 km s$^{-1}$ of stellar rotation. For stars rotating faster than this, further increasing $R$ will thus provide a small change in the measured velocity uncertainty. Generically, the limiting spectral resolution for a given stellar rotation velocity will be 
\begin{equation}\label{eq:6010}
\frac{R_{lim}}{\mathrm{60,000}} = \frac{4.25\ \mathrm{km\,s}^{-1}}{v_{rot}}. 
\end{equation}
That being said, extremely high spectral resolutions may help with correcting for systematics in observations, by enabling one to study the detailed shapes of the lines. 

\begin{figure}[t]
\vskip -0.0in 
\epsscale{1.2} 
\plotone{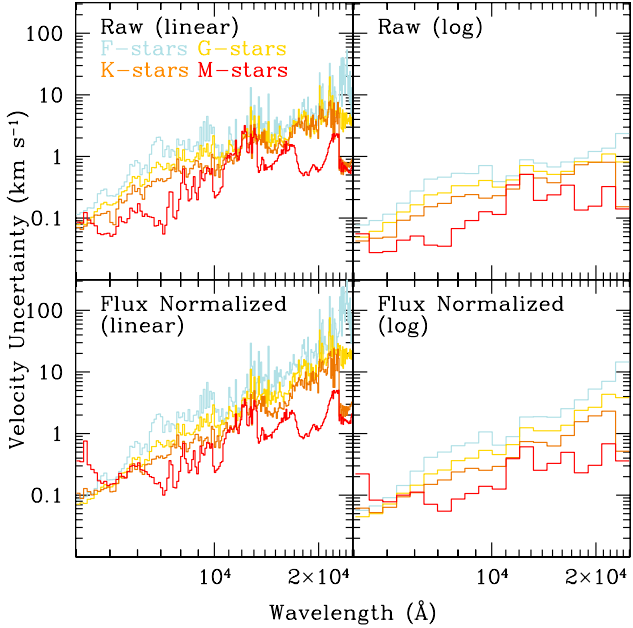}
\vskip -0.0in
\caption{The base uncertainty values from Table 1 divided into M-stars (2600K to 4000K), K-stars (4000K to 5200K), G-stars (5200K to 6000K), and F-stars (6000K to 7600K) and median combined within those categories. The two left panels show the medians for each 100\,\AA\ wavelength chunk in Table 1, while the right panels show the median values after the chunks have been organized in 20 logarithmically-spaced bins. This is meant to replicate a spectrograph observing at fixed resolution. The top panels show the raw numbers from Table 1, while the bottom panels are normalized to according to the fraction of the overall bolometric luminosity that each chunk occupies.}
\label{fig:wavemedian}
\end{figure}

Third, Figures \ref{fig:optq}, \ref{fig:redq}, and \ref{fig:nirq} show the importance of choosing the appropriate wavelength for an RV survey, particularly if one cannot cover a wide range of wavelengths. To investigate this in more detail, we collapsed the uncertainty values behind Figures  \ref{fig:optq}, \ref{fig:redq}, \ref{fig:nirq}, and Table 1 along the temperature axis to see what are the best locations for observations. We divided the temperature range into M-stars (2600K to 4000K), K-stars (4000K to 5200K), G-stars (5200K to 6000K), and F-stars (6000K to 7600K) and took the median uncertainty values across these ranges for each 100\,\AA\ wavelength chunk in Table 1. Figure \ref{fig:wavemedian} shows the results, visualized in four different ways. First, the upper left panel shows the median uncertainties of each 100\,\AA\ chunk for the four spectral types (labeled ``Raw (linear)''). The lower left panel shows these same chunk medians, but now we have normalized them according to the fraction of the overall blackbody luminosity that each chunk occupies (labeled ``BB Normalized (linear)''). In terms of Figures  \ref{fig:erropt}, \ref{fig:errred}, and \ref{fig:errnir}, this factors in the ``BB'' line. The two right panels in Figure \ref{fig:wavemedian} are similarly ``Raw'' or ``BB Normalized,'' but we have now combined the chunks into equally wide logarithmic bins, rather than plotting them linearly as before. This is meant to replicate the true observing mode of spectrographs: at fixed resolution an instrument can image proportionally more of a spectrum at proportionally longer wavelengths. The lower right panel of Figure \ref{fig:wavemedian} therefore most directly informs the selection of a proper observing wavelength.

In the optical, spectra longwards of 5500\,\AA\ provide relatively little information for F-stars, and provide an inferior amount of velocity information for later spectral types as compared to most of the shorter wavelengths. For a survey covering all spectral classes with a limited wavelength range, the best wavelengths to look at in the optical would be from 5000\,\AA\ to 5200\,\AA, which contains the Mg b triplet, and has been noted previously \citep{latham1985}. Shorter wavelengths than this, while advantageous for FG stars, provide little gain for either K- or M-stars.

In the ``red,'' between 6500\,\AA\ and 10000\,\AA, the best regions for a general exoplanet survey are in the I or $i'$ bands around 7000\,\AA. Moving further out into the red, to $z'$, provides less velocity information at almost all effective temperatures. Indeed, as shown in the bottom right panel of Figure \ref{fig:wavemedian}, observing from 7000\,\AA\ to 8000\,\AA\ in I or $i'$ has the lowest base uncertainty for observations of M-dwarfs over the entire wavelength range.

\begin{figure}[t]
\vskip -0.0in 
\epsscale{1.2} 
\plotone{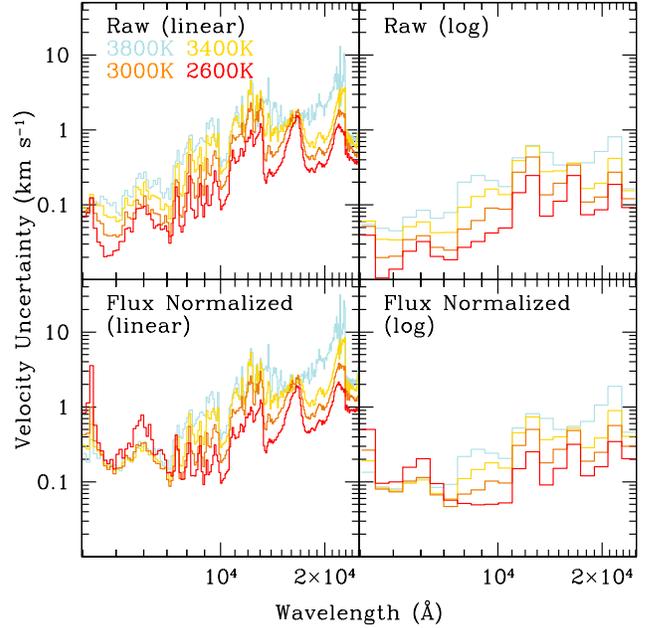}
\vskip -0.0in
\caption{The base uncertainty values from Table 1 for four different M-dwarf temperatures. As in Figure \ref{fig:wavemedian}, the two left panels show the uncertainties for the 100\,\AA\ wavelength chunks listed in Table 1, while the right panels show the uncertainties after the chunks have been combined into 20 logarithmically-spaced bins. This is meant to replicate a spectrograph observing at fixed resolution. The top panels show the raw numbers from Table 1, while the bottom panels are normalized to according to the fraction of the overall bolometric luminosity that each chunk occupies. The bottom right panel particularly illustrates that the most efficient wavelength range to observe M-dwarfs is generally shortwards of 10,000\,\AA.}
\label{fig:wavemedianmdwarfs}
\end{figure}

To investigate this in more detail, Figure \ref{fig:wavemedianmdwarfs} shows four of the temperatures that make up the M-dwarfs in Figure \ref{fig:wavemedian} displayed in a similar manner. One can see that between 3800K (an M0) and 2600K (roughly an M7) the shape of the raw uncertainties as a function of wavelength are roughly self-similar, but the flux normalized uncertainties flatten out as one moves to cooler temperatures. This is, of course, a result of the peak of the stellar SED moving towards longer wavelengths as the temperature decreases. Nevertheless, although the peak of a 2600K blackbody has moved out to 11,000\,\AA, one can clearly see in the lower panels of Figure \ref{fig:wavemedianmdwarfs} that the most efficient wavelength region to observe all of the M-dwarf temperatures we consider is shortwards of 10,000\,\AA, between 6000\,\AA\ and 9000\,\AA. The M-dwarf spectra at these wavelengths have approximately 10 times the velocity information (i.e., 1/10 the uncertainty) as compared to the NIR, and thus remain the most efficient observing location even after accounting for the relative amount of stellar emission. \cite{bottom2013} arrived at a similar conclusion for the best wavelength regime to observe M-dwarfs. This illustrates the importance of considering the velocity information available in a wavelength region, and not just the shape of the stellar SED, when designing an RV survey targeted at M-dwarfs. 

If we now consider only the NIR, it is interesting to compare Figures \ref{fig:nirq} and \ref{fig:wavemedian} against the available transmission windows in the atmosphere. We can roughly approximate these windows by considering the NIR photometry bands: \emph{J} (11000\,\AA\ to 13500\,\AA), \emph{H} (15000\,\AA\ to 17000\,\AA), and \emph{K} (20000\,\AA\ to 23000\,\AA). For a general RV survey of all spectral types, the optimum observing band would be in \emph{J}, since the uncertainty for FGK spectral types steadily increases towards longer wavelengths. For a survey targeting only later spectral types, the decision is less clear cut. \emph{H}-band has the lowest uncertainty for observing M-dwarfs (down 10\% compared to \emph{J} and down 40\% compared to \emph{K}). At the same time, if one wished to include K-dwarfs in this survey then the optimum observing bandpass would again be \emph{J}. It is interesting to note that the best wavelength regions to observe M-dwarfs are precisely in between \emph{J}, \emph{H}, and \emph{K}. This is a result of molecular lines from water appearing in the cooler stars' atmospheres, but this same water absorption in Earth's atmosphere is precisely what sets the location of the NIR observing bands.

\section{Summary}

We have considered the astrophysical sources of velocity uncertainty in stellar RV measurements. In doing so, we are able to describe the basic mechanisms that cause velocity uncertainties, what the dominant driver is behind stellar velocity uncertainties at various stellar masses and in various wavelength regions, and furnish several points for consideration when designing an RV survey. In doing so, we emphasize that we have deliberately decided to restrict our focus to astrophysical sources of uncertainty, and so we do not consider the effects of star spots, granulation, or asteroseismic pulsations. Similarly, we do not treat instrumental velocity uncertainties like wavelength calibration, optical effects, or instrumental drifts, nor do we consider the effect of telluric absorption lines on the spectra.  

We determine general scaling laws for the expected velocity uncertainty. This allows the reader to estimate the amount of velocity uncertainty present in observations of main-sequence stars using an arbitrary wavelength range between 4000\,\AA\ and 25000\,\AA, over a large set of possible spectral resolutions and stellar properties. This is in contrast to previous work in this area, which has provided results using specific observing set-ups (e.g., fixed spectral resolution or stellar properties). 

At a basic level, we demonstrate that the velocity uncertainty of a weak spectral absorption line in a continuum scales as $\Theta^{3/2}$, where $\Theta$ is the FWHM of the line, and not linearly with $\Theta$ as one expects when there is no continuum emission and as has been claimed in some previous work. Using model spectra, we then calculated how the velocity uncertainty changes as a function of spectral resolution, stellar rotation, stellar effective temperature, stellar surface gravity, and stellar metallicity. By dividing our model spectra up into 100\,\AA-wide chunks, we find that the effects of resolution, rotation, and surface gravity operate on the chunks in a largely self-similar manner -- regardless of the specific wavelength or spectral features within a chunk. Effective temperature presents a more complicated picture, with different chunks behaving very differently. We numerically fit a rough relation to the chunk medians, but the variation between chunks as a function of temperature is one illustration of the importance of carefully choosing the wavelength range used in an RV survey.

With these basic relations established, we are able to calculate how the velocity uncertainty scales as a function of stellar mass. For stars more massive than 1.1$M_\odot$, we find that the rapidly increasing stellar rotation dominates the predicted uncertainties. Below 1.1$M_\odot$, the velocity uncertainty is set by a combination of competing effects from changes in stellar luminosity, temperature and surface gravity. In the optical, between 4000\,\AA\ and 6500\,\AA, we find that almost all of these effects cancel, leaving the velocity uncertainty to be predominately set by the bolometric luminosity of the target star for a fixed distance. This is not true in the red (6500\,\AA\ to 10000\,\AA) or the NIR (10000\,\AA\ to 25000\,\AA), where one must also account for temperature (red) or temperature and the effect of the blackbody peak shifting relative to the observed wavelength range (NIR). We give simple approximations for how the velocity error scales with mass for each of these three wavelength regimes.  

More generally, our consideration of velocity errors in RV surveys highlights two important points for consideration. First, after a certain point increasing spectral resolution provides diminished returns. This primary occurs because once one has resolved the lines in a spectrum increased resolution provides little more information, and because, depending on the stars being surveyed, stellar rotation will provide the dominant source of velocity uncertainty -- not spectral resolution. We find that this point of diminishing returns occurs at approximately $R$=80,000, though we note that extremely high spectral resolutions may help with correcting for systematics in observations.

Second, the most efficient wavelength region to operate an RV survey for M-dwarfs is between 6000\,\AA\ to 9000\,\AA. Although the peak emission for M-dwarfs is generally longwards of these wavelengths, the base velocity uncertainties of spectra in this wavelength region are about 1/10 that of spectra in the NIR bands. This means that even after accounting for the difference in received flux, M-dwarf spectra from 6000\,\AA\ to 9000\,\AA\ will give a lower velocity uncertainty than spectra in observed in the NIR at the same exposure time.

\acknowledgments
We would like to thank the anonymous referee for their help in improving the manuscript. We would also like to thank Jason Wright for his thoughts and comments.

This work was partially supported by NSF CAREER Grant AST-105652.

\end{document}